\documentclass[journal]{IEEEtran}
\ifCLASSINFOpdf
\else
\fi
\usepackage{amsmath}
\usepackage{amsmath,amssymb}
\usepackage[mathscr]{euscript}
\usepackage{algorithmic}
\usepackage[ruled]{algorithm2e}

\usepackage{array}
\usepackage{graphicx}
\usepackage[caption=false,font=normalsize,labelfont=sf,textfont=sf]{subfig}
\usepackage{url}
\begin{document}
\title{Data-Driven Model Reduction for Multilinear Control Systems via Tensor Trains}

\author{Can Chen, Amit Surana,~\IEEEmembership{Member,~IEEE}, Anthony Bloch,~\IEEEmembership{Fellow,~IEEE}, Indika Rajapakse
\thanks{C. Chen is with Department of Mathematics and Department of Electrical Engineering and Computer Science, University of Michigan,
    {\tt \small canc@umich.edu}}
\thanks{A. Surana is with the United Technologies Research Center,
        411 Silver Lane, East Hartford, CT 06108,
        {\tt \small suranaa@utrc.utc.com}} %
\thanks{A. Bloch is with Department of Mathematics, University of
    Michigan {\tt \small abloch@umich.edu}}
\thanks{I. Rajapakse is with the Department of Computational Medicine \& Bioinformatics, Medical School and Department of Mathematics, University of Michigan, {\tt \small indikar@umich.edu}} %

\thanks{Manuscript received \today.}}


\markboth{IEEE TRANSACTIONS ON CONTROL SYSTEMS TECHNOLOGY,~Vol.~X, No.~X, X~20XX}%
{Chen \MakeLowercase{\textit{et al.}}: Data-Driven Model Reduction for Multilinear Control Systems via Tensor Trains}



\maketitle

\begin{abstract}
In this paper, we explore the role of tensor algebra in balanced truncation (BT) based model reduction/identification for high-dimensional multilinear/linear time invariant systems. In particular, we employ tensor train decomposition (TTD), which provides a good compromise between numerical stability and level of compression, and has an associated algebra that facilitates computations. Using TTD, we propose a new BT approach which we refer to as higher-order balanced truncation, and consider different data-driven variations including higher-order empirical gramians, higher-order balanced proper orthogonal decomposition and a higher-order eigensystem realization algorithm. We perform computational and memory complexity analysis for these different flavors of TTD based BT methods, and compare with the corresponding standard BT methods in order to develop insights into where the proposed framework may be beneficial. We provide numerical results on simulated and experimental datasets showing the efficacy of the proposed framework.
\end{abstract}


% \fi
\IEEEpeerreviewmaketitle

\section{Introduction}
\IEEEPARstart{T}{he} goal of this paper is to explore the role of tensor algebra in data-driven model reduction/identification for high-dimensional multilinear/linear time invariant (MLTI/LTI) input-output systems. Tensors are multidimensional arrays generalized from vectors and matrices, and have wide applications in many domains such as social networks, biology, cognitive science, applied mechanics, scientific computations and signal processing \cite{doi:10.1137/100804577,Chen_2019,tensorSPbook,HoKoDu19,Kolda06multilinearoperators,kolda2009tensor,WILLIAMS20181099}. Tensor representation preserves multidimensional patterns, capturing higher-order interactions and couplings within multiway data, instead of the standard pairwise ``flattened" view inherent in two-way matrix based analysis. Tensor decomposition techniques such as CANDECOMP/PARAFAC decomposition (CPD) \cite{doi:10.1137/060676489,Kolda06multilinearoperators,kolda2009tensor}, higher-order singular value decomposition (HOSVD) \cite{doi:10.1137/S0895479896305696} and tensor train decomposition (TTD) \cite{doi:10.1137/090752286,doi:10.1137/090748330} help reveal such hidden patterns/redundancies to obtain a compact representation, thereby reducing storage effort and enabling efficient computations.

Tensor algebra has been recently exploited in systems and control applications. The key idea is to tensorize the vector based dynamic system representation into an equivalent tensor representation, and to exploit tensor algebra. A large body of literature, referred to as tensor product models, has emerged which utilizes compact tensor based representations/computations in the context of linear parameter varying models \cite{baranyi2013tensor}. Other applications include efficient solution of Lyapunov equations \cite{6760167}, fault detection \cite{muller2015fault}, Kalman filtering \cite{batselier2017tensor}, accelerating simulation of nonlinear models where the vector field is a multilinear function of states \cite{kruppa2017comparison}, and modeling inverse dynamics \cite{baier2017tensor}, to name a few.

In many scientific and engineering applications, the system dynamics over space and time is often described in terms of partial differential equations (PDEs), e.g., Navier Stokes and heat equations in thermal/fluids, Euler equations in structural mechanics, Schr{\"o}dinger equations in quantum mechanics, etc. The system state in such representation is a field (e.g., temperature/velocity field) whose discretization in space/time naturally results in tensors evolving with time. In order to apply the standard model reduction/identification framework, such as Balanced Truncation (BT) \cite{doi:10.1142/S0218127405012429}, Eigensystem Realization Algorithm (ERA) \cite{doi:10.2514/3.20031,Ma2011} or Dynamic Mode Decomposition (DMD) and its variants \cite{doi:10.1137/15M1013857,tu_2014}, tensors need to be vectorized. This may result in an extremely high-dimensional system representation in which the number of states/model parameters scales exponentially with the number of dimensions of the tensors involved, and thus pose a significant computational challenge. Alternatively, a new class of MLTI systems has been introduced \cite{doi:10.1137/1.9781611975758.18, Chen_2019,rogers_2013,7798500} in which the states and outputs are preserved as tensors, and the system evolution is generated by the action of multilinear operators. By using tensor unfolding, an operation that transforms a tensor into a matrix, Rogers et al. \cite{rogers_2013} and Surana et al. \cite{7798500} developed methods for model reduction/identification from tensor time series data, and demonstrated benefits such as a more compact and accurate representation compared to the classical vectorization based LTI approach. An application of such tensor based representation and identification for skeleton based human behavior recognition from videos demonstrated significant improvements in classification accuracy compared to standard LTI based approaches \cite{ding2018tensor}. 
%The TTD framework has been applied to accelerate DMD computations \cite{Klus_2018}, and for computing numerical solutions of master equations associated with high-dimensional Markov processes \cite{gel2017tensor}.

Chen et al. \cite{doi:10.1137/1.9781611975758.18,Chen_2019} generalized the notion of MTLI systems to incorporate control inputs based on the Einstein product and even-order paired tensors. By leveraging recent advances in tensor algebra, Chen et al. also developed tensor generalizations of classical LTI system notions including stability, reachability and observability, and expressed them in different forms including concepts based on tensor unfolding and other more standard notions of tensor ranks/decompositions \cite{Chen_2019}. In addition, the authors in \cite{Chen_2019} demonstrated that the generalized CPD and TTD (GTTD) based model reduction framework can significantly reduce the number of MLTI system parameters. Continuing along that line of work, in this paper, we develop a novel tensor based BT computational framework for model reduction/identification in high-dimensional MLTI/LTI systems. In particular, we choose to work with TTD as the underlying computational framework since it provides a good compromise between numerical stability and level of compression. The TTD framework has been applied to accelerate DMD computations \cite{Klus_2018}, and for computing numerical solutions of master equations associated with high-dimensional Markov processes \cite{gel2017tensor}.
%In addition TTD has an associated algebra where basic operations such as addition, scalar product, matrix-by-vector product and norms etc. can be done and maintained in the TT format, without needing to go to the full tensor representation.
Moreover, the proposed framework is naturally suited for MLTI systems, but can also be applied to standard vector based LTI system representations by tensorizing them into a suitable tensor form \cite{6760167,doi:10.1137/110833142}. The key contributions of this paper are as follows:
\begin{itemize}
\item We develop a TTD based computational framework for different flavors of BT: Higher-Order Balanced Truncation (HOBT) which involves solutions of tensor Lyapunov equations to obtain reachability and observability gramians, and subsequent computations of the balancing transform; Higher-Order Empirical Gramians (HOEG) which constructs reachability and observability gramians from snapshots via simulation of forward/adjoint MTLI system equations; Higher-Order Balanced Proper Orthogonal Decomposition (HOBPOD) which directly and more efficiently constructs the balancing transform from forward/adjoint snapshots; and Higher-Order Eigensystem Realization Algorithm (HOERA) which is equivalent to HOBPOD but only requires snapshots from the forward model, and so can also be applied to experimental data.
\item We employ TT-algebra for the memory efficient computations whereby basic tensor operations such as addition, scalar product, Einstein product, norms, solution to multilinear equations and tensor pseudoinverse, can be computed and maintained in the TT-format, without needing to go to the full tensor representation. We also present variations for computing block tensors and tensor singular value decomposition (TSVD) using TT-algebra, which are the key operations in HOBPOD/HOERA.
\item We provide computational and memory complexity analysis for different tensor based BT methods, and compare them with the corresponding standard BT methods. This analysis provides insights into when the tensor based representation can be beneficial, and also suggests a framework whereby one can mix tensor and matrix approaches to gain the best efficiency.
\item We demonstrate our framework in four numerical examples including a synthetic dataset, the 2D heat equation with control, a cancer cell image video dataset and a room impulse responses dataset. A comparison with the standard BT is provided indicating the memory and computational time savings.
\end{itemize}

The paper is organized into six sections. We start with the basics of tensor algebra followed by descriptions of even-order paired tensors, tensor decompositions and TT-algebra in section \ref{sec:2}. We also establish results about the construction of block tensors and the computation of TSVD using TTD. In section \ref{sec:3}, we present a generalization of the balanced truncation, balanced proper orthogonal decomposition and eigensystem realization algorithm framework for MLTI systems with detailed algorithms and complexity analysis. Four numerical examples are presented in section \ref{sec:4}. Finally, we discuss some directions for future research in section \ref{sec:5} and conclude in section \ref{sec:6}. For ease of reading, we provide a list of acronyms in Appendix \ref{apd:B}.

\section{Preliminaries}\label{sec:2}
In this section, we briefly review tensor preliminaries. Comprehensive reviews can be found in \cite{Chen_2019,gel2017tensor,Kolda06multilinearoperators,kolda2009tensor,doi:10.1137/090752286}. An $N$-th order \textit{tensor} usually is denoted by $\textsf{X}\in \mathbb{R}^{J_1\times J_2\times  \dots \times J_N}$. The sets of indexed indices and size of $\textsf{X}$ are denoted by $\textbf{j}=\{j_1,j_2,\dots,j_N\}$ and $\mathcal{J}=\{J_1,J_2,\dots,J_N\}$, respectively. $|\mathcal{J}|$ represents the product of all elements in $\mathcal{J}$, and $\textbf{j}\in[\mathcal{J}]$ can be interpreted as $j_n=1,2,\dots,J_n$ for $n=1,2,\dots,N$. For more compact notation, we define $\textsf{X}_\textbf{j} = \textsf{X}_{j_1j_2\dots j_N}$ with $\mathbb{R}^{\mathcal{J}} = \mathbb{R}^{J_1\times J_2\times  \dots \times J_N}$ and $\textsf{X}_{\textbf{j}\textbf{i}} = \textsf{X}_{j_1\dots j_Ni_1\dots i_N}$ with $\mathbb{R}^{\mathcal{J}\times \mathcal{I}} = \mathbb{R}^{J_1\times \dots\times J_N\times I_1\times  \dots \times I_N}$. The \textit{tensor inner product} of two tensors of the same size is defined by
$\langle \textsf{X},\textsf{Y}\rangle =\sum_{\textbf{j}=\textbf{1}}^{\mathcal{J}}\textsf{X}_{\textbf{j}}\textsf{Y}_{\textbf{j}}
$
where the notation $\sum_{\textbf{j}=\textbf{1}}^{\mathcal{J}}$ is an abbreviation of the $N$ summations over all indices $\textbf{j}\in[\mathcal{J}]$.  The \textit{tensor Frobenius norm} induced by the inner product is given as $\|\textsf{X}\|^2=\langle \textsf{X},\textsf{X}\rangle$. The \textit{matrix tensor multiplication} $\textsf{X} \times_{n} \textbf{A}$ along mode $n$ for a matrix $\textbf{A}\in  \mathbb{R}^{I\times J_n}$ is defined by
$
(\textsf{X} \times_{n} \textbf{A})_{j_1j_2\dots j_{n-1}ij_{n+1}\dots j_N}=\sum_{j_n=1}^{J_n}\textsf{X}_{\textbf{j}}\textbf{A}_{ij_n}.
$
This product can be generalized to what is known as the \textit{Tucker product}, for $\textbf{A}_n\in \mathbb{R}^{I_n\times J_n}$,
\begin{equation*}
\begin{split}
    &\textsf{X}\times_1 \textbf{A}_1 \times_2\textbf{A}_2\times_3\dots \times_{N}\textbf{A}_N\\=&\textsf{X}\times\{\textbf{A}_1,\textbf{A}_2,\dots,\textbf{A}_N\}\in  \mathbb{R}^{\mathcal{I}}.
\end{split}
\end{equation*}

\subsection{Even-order paired tensors}
The notion of \textit{even-order paired tensors} was first proposed by Huang et al. \cite{Huang_2017} in the context of solid mechanics. For an arbitrary even-order tensor $\textsf{A}\in\mathbb{R}^{J_1\times I_1\times \dots \times J_N\times I_N}$, if its indices can be divided into $N$ adjacent blocks $\{j_1i_1\},\dots, \{j_Ni_N\}$, then $\textsf{A}$ is called an even-order paired tensor \cite{doi:10.1137/1.9781611975758.18}. Similarly, we define $\textsf{A}_{\textbf{j}\otimes\textbf{i}} = \textsf{A}_{j_1i_1\dots j_Ni_N}$ with $\mathbb{R}^{\mathcal{J}\otimes \mathcal{I}} = \mathbb{R}^{J_1\times I_1\times \dots \times J_N\times I_N}$ for simplicity. Building on the work by Brazell et al. \cite{doi:10.1137/100804577}, Chen et al. \cite{doi:10.1137/1.9781611975758.18} proposed an unfolding transformation $\psi$ from the even-order paired tensor space $\mathbb{T}_{\mathcal{J}\otimes\mathcal{I}}(\mathbb{R})$ to the matrix space $\mathbb{M}_{|\mathcal{J}||\mathcal{I}|}(\mathbb{R})$ defined by
\begin{equation}
\textsf{A}_{\textbf{j}\otimes\textbf{i}}\xrightarrow{\psi} \textbf{A}_{[ j_1+\sum_{k=2}^N(j_k-1)\prod_{l=1}^{k-1}J_l][i_1+\sum_{k=2}^N(i_k-1)\prod_{l=1}^{k-1}I_l]},
\end{equation}
and showed that $\psi$ is a ring isomorphism for $\mathcal{J}=\mathcal{I}$ under the element-wise addition and the \textit{Einstein product} defined by
\begin{equation}\label{eq2}
(\textsf{A}*\textsf{B})_{\textbf{j}\otimes\textbf{i}}= \sum_{\textbf{k}=\textbf{1}}^{\mathcal{K}}\textsf{A}_{\textbf{j}\otimes\textbf{k}}\textsf{B}_{\textbf{k}\otimes\textbf{i}},
\end{equation}
for $\textsf{A}\in\mathbb{R}^{\mathcal{J}\otimes \mathcal{K}}$ and $\textsf{B}\in\mathbb{R}^{\mathcal{K}\otimes \mathcal{I}}$. Based on the transformation $\psi$, one can define matrix-like notions for tensor algebra including \textit{U-transpose}, \textit{U-diagonal}, \textit{U-identity}, \textit{U-orthogonal}, \textit{U-inverse}, \textit{U-positive definiteness} and \textit{U-eigenvalues} (see \cite{Chen_2019} for details). Liang et al. \cite{doi:10.1080/03081087.2018.1500993} also define the \textit{unfolding rank} of an even-order paired tensor $\textsf{A} \in \mathbb{R}^{\mathcal{J}\otimes \mathcal{I}}$ by  $\text{rank}_U(\textsf{A}) = \text{rank}\big(\psi(\textsf{A})\big)$.

Chen et al. \cite{doi:10.1137/1.9781611975758.18} introduce notions of block tensors for same-size even-order paired tensors. For  $\textsf{A},\textsf{B}\in\mathbb{R}^{\mathcal{J}\otimes \mathcal{I}}$, the \textit{$n$-mode row block tensor} is defined to be $\begin{vmatrix}
	\textsf{A} & \textsf{B}
\end{vmatrix}_n\in \mathbb{R}^{\mathcal{J}\otimes \mathcal{L}}$ such that
\begin{equation}
(\begin{vmatrix}
	\textsf{A} & \textsf{B}
\end{vmatrix}_n)_{\textbf{j}\otimes\textbf{l}}=
\begin{cases}
\textsf{A}_{\textbf{j}\otimes\textbf{l}}, \text{ }\textbf{j}\in[\mathcal{J}], \text{ } \textbf{l}\in[\mathcal{I}]\\
\textsf{B}_{\textbf{j}\otimes\textbf{l}}, \text{ }\textbf{j}\in[\mathcal{J}], \text{ } \textbf{l}\in[\mathcal{L}]
\end{cases},
\end{equation}
where, $\mathcal{L} = \{I_1,I_2,\dots,2I_n,\dots,I_N\}$, and $\textbf{l}\in[\mathcal{L}]$ represents $l_n=I_n+1,I_n+2,\dots,2I_n$. The block tensor $\begin{vmatrix}\textsf{A} & \textsf{B} \end{vmatrix}_n$ is simply the concatenation of \textsf{A} and \textsf{B} at the $2n$-th mode, which can be generalized to an arbitrary number of even-order paired tensors. More generally, given $K$ even-order paired tensors $\textsf{X}_k\in\mathbb{R}^{\mathcal{J}\otimes \mathcal{I}}$ and a factorization $K=K_1K_2\dots K_N$, the $\mathcal{J}\otimes \mathcal{I}\mathcal{K}$ order \textit{mode row block tensor} $\textsf{Y}$ can be constructed in the following way:
 first, compute the 1-mode row block tensor concatenation over  $\{\textsf{X}_1,\cdots,\textsf{X}_{K_1}\}$, $\{\textsf{X}_{K_1+1},\cdots,\textsf{X}_{2K_1}\}$ and so on to obtain $K_2K_3\dots K_N$ block tensors denoted by $\textsf{X}_1^{(1)},\textsf{X}_2^{(1)},\dots,\textsf{X}_{K_2K_3\dots K_N}^{(1)}$;
 second, compute the 2-mode row block tensors concatenation over $\{\textsf{X}_1^{(1)},\cdots,\textsf{X}_{K_2}^{(1)}\}$, $\{\textsf{X}_{K_2+1}^{(1)},\cdots,\textsf{X}_{2K_2}^{(1)}\}$ and so on to obtain $K_3K_4\dots K_N$ block tensors denoted by $\textsf{X}_1^{(2)},\textsf{X}_2^{(2)},\dots,  \textsf{X}_{K_3K_4\dots K_N}^{(2)}$;
 third, keep repeating the process until the last $N$-mode row block tensor is obtained which we denote by
 $\textsf{Y} = \begin{vmatrix} \textsf{X}_1 &\textsf{X}_2 & \dots & \textsf{X}_{K}\end{vmatrix}$,
where $\mathcal{I}\mathcal{K}=\{I_1K_1,I_2K_2,\dots,I_NK_N\}$.

Analogously, one can also define the notions of \textit{$n$-mode column block tensors} and \textit{mode column block tensors}. Mode row/column block tensors possess many useful matrix-like properties such as the block tensor Einstein product which are useful in MLTI systems theory \cite{Chen_2019}. Moreover, the blocks of mode row/column block tensors map to contiguous blocks under $\psi$ up to some permutations \cite{Chen_2019, doi:10.1137/110820609}. One can also define the inverse operation of extracting the component tensors from a given mode row block tensor. Let $\textsf{Y}\in\mathbb{R}^{\mathcal{J}\otimes \mathcal{I}\mathcal{K}}$ be a mode row block tensor constructed using the factorization $K=K_1K_2\dots K_N$, then the component tensors $\textsf{X}_k\in\mathbb{R}^{\mathcal{J}\otimes \mathcal{I}},k=1,2,\dots,K,$ can be extracted as:
\begin{equation}\label{eq:inverseblock}
\textsf{X}_k = \textsf{Y}_{:[k_1I_1-I_1+1:k_1I_1]\dots:[k_NI_N-I_N+1:k_NI_N]},
\end{equation}
where, $k=k_1+\sum_{i=2}^N(k_i-1)\prod_{l=1}^{i-1}K_l$ for $k_n=1,2,\dots,K_n$. The colon operation : in (\ref{eq:inverseblock}) is a way to refer to the slices of a tensor as used in MATLAB.

\subsection{Tensor decompositions}
There are a variety of notions of tensor decompositions such as CANDECOMP/PARAFAC decomposition, higher-order singular value decomposition, Tucker decomposition, tensor train decomposition and tensor singular value decomposition, which all play an important role in tensor algebra \cite{doi:10.1137/100804577, Chen_2019, doi:10.1137/S0895479896305696, gel2017tensor,Kolda06multilinearoperators,kolda2009tensor,doi:10.1137/090752286, doi:10.1137/090748330}. Of particular interest in this paper are the tensor train decomposition (TTD) and the tensor singular value decomposition (TSVD).

The TTD of an $N$-th order tensor $\textsf{X}\in\mathbb{R}^{\mathcal{J}}$ is given by
\begin{equation}\label{train}
\textsf{X} = \sum_{\textbf{r} = \textbf{1}}^{\mathcal{R}} \textsf{X}^{(1)}_{r_0:r_1} \circ \textsf{X}^{(2)}_{r_1:r_2}\circ \dots \circ \textsf{X}^{(N)}_{r_{N-1}:r_N},
\end{equation}
where, $\circ$ is the outer product, $\mathcal{R} = \{R_0,R_1,\dots,R_N\}$ is the set of TT-ranks with $R_0=R_N=1$, and $\textsf{X}^{(n)}\in\mathbb{R}^{R_{n-1}\times J_n\times R_{n}}$ are called the core tensors of the TTD. There exist optimal TT-ranks such that
$$R_n=\text{rank}\big(\texttt{reshape}(\textsf{X},\prod_{i=1}^nJ_i,\prod_{i=n+1}^N J_i)\big),$$ 
for $n=1,2,\dots,N-1$ \cite{doi:10.1137/090752286}. A core tensor $\textsf{X}^{(n)}$ is called \textit{left-orthonormal} if $(\bar{\textbf{X}}^{(n)})^\top\bar{\textbf{X}}^{(n)} = \textbf{I}\in\mathbb{R}^{R_n\times R_n}$, and is called \textit{right-orthonormal} if $\underline{\textbf{X}}^{(n)}(\underline{\textbf{X}}^{(n)})^\top = \textbf{I}\in\mathbb{R}^{R_{n-1}\times R_{n-1}}$ where
$
\bar{\textbf{X}}^{(n)} = \texttt{reshape}(\textsf{X}^{(n)},R_{n-1}J_n,R_n)$ and $
\underline{\textbf{X}}^{(n)} = \texttt{reshape}(\textsf{X}^{(n)},R_{n-1},J_nR_n)
$
are the left- and right-unfoldings of the core tensor, respectively \cite{Klus_2018}. Here \textbf{I} denotes the identity matrix, and \texttt{reshape} refers to the reshape operation in MATLAB. Truncating the TT-ranks results in a quasi-optimal approximation of $\textsf{X}$ \cite{doi:10.1137/090752286}.

One can also define tensor trains for even-order paired tensors. Given an even-order paired tensor $\textsf{A}\in\mathbb{R}^{\mathcal{J}\otimes \mathcal{I}}$, the generalized TTD (GTTD) of  $\textsf{A}$ is defined by
\begin{equation}\label{eq:7}
\textsf{A}=\sum_{\textbf{r}=\textbf{1}}^\mathcal{R}\textsf{A}^{(1)}_{r_0::r_1}\circ \textsf{A}^{(2)}_{r_1::r_2}\circ \dots \circ \textsf{A}^{(N)}_{r_{N-1}::r_N},
\end{equation}
where, $\textsf{A}^{(n)}\in\mathbb{R}^{R_{n-1}\times J_n\times I_n\times R_n}$, and $\mathcal{R}$ is the set of GTT-ranks with $R_0=R_N=1$ \cite{Chen_2019,gel2017tensor, doi:10.1137/110833142}. Clearly, TTD is a special case of  GTTD with $\mathcal{I} = \textbf{1}$. One may even quantize the tensor trains (\ref{eq:7}) at each dimension, e.g., $J_n = J_{n1}J_{n2}\dots J_{nm_n}$ for some positive integer $m_n$, in which a typical choice of $J_{nm_n}$ is 2,  in order to further reduce the complexity of  TTD/GTTD and accelerate computations if the  QTT-ranks are small  \cite{Khoromskij2011,Oseledets2009,doi:10.1137/090757861}. We will refer to it as quantized TTD (QTTD). In fact, the QTTD of $\textsf{A}\in\mathbb{R}^{\mathcal{J}\otimes\mathcal{I}}$ is the GTTD of the reshaped tensor $\tilde{\textsf{A}}\in\mathbb{R}^{(\mathcal{J}_1\otimes \mathcal{I}_1)\times \dots \times (\mathcal{J}_N\otimes \mathcal{I}_N)}$ where $\mathcal{J}_n = \{J_{n1},J_{n2},\dots,J_{nm_n}\}$ and $\mathcal{I}_n = \{I_{n1},I_{n2},\dots,I_{nm_n}\}$ such that $|\mathcal{J}_n|=J_n$ and $|\mathcal{I}_n|=I_n$, respectively. A detailed algorithm for the conversion can be found in \cite{gel2017tensor}. Conversely, we define an operation $\zeta(\cdot)$ that can recover full tensor format from its TTD/GTTD/QTTD. 

TSVD was first proposed by Brazell et al. \cite{doi:10.1137/100804577} based on the isomorphism property and was extended by Sun et al. \cite{doi:10.1080/03081087.2015.1083933} for general non-paired even-order tensors. The results can be easily extended to even-order paired tensors. The economy-size TSVD (ETSVD) of an even-order paired tensor $\textsf{A}\in\mathbb{R}^{\mathcal{J}\otimes \mathcal{I}}$ can be written as
\begin{equation}\label{eq33}
\textsf{A} = \textsf{U}*\textsf{S}*\textsf{V}^{\top},
\end{equation}
where, the superscript $\top$ denotes the U-transpose operation (see next subsection), $\textsf{U}\in \mathbb{R}^{\mathcal{J}\otimes \mathcal{R}}$ and $\textsf{V}\in \mathbb{R}^{\mathcal{I}\otimes \mathcal{R}}$ such that $\textsf{U}^\top*\textsf{U} = \textsf{I}$ and $\textsf{V}^\top*\textsf{V} = \textsf{I}$ (\textsf{I} denotes the U-identity tensor), $\textsf{S}\in \mathbb{R}^{\mathcal{R}\otimes \mathcal{R}}$ is an U-diagonal tensor containing the singular values of \textsf{A} along its diagonal $\textsf{S}_{\textbf{r}\otimes \textbf{r}}$, and $\mathcal{R}=\{R_1,R_2,\dots,R_N\}$ such that $|\mathcal{R}|$ is equal to the unfolding rank of \textsf{A}. The ETSVD (\ref{eq33}) also can be rewritten as
\begin{equation}
\textsf{A}=\sum_{r=1}^{|\mathcal{R}|}\sigma_{r} \textsf{X}_{r}* \textsf{Y}_{r}^{\top},\label{eq34}
\end{equation}
where, $\textsf{X}_{r}\in\mathbb{R}^{\mathcal{J}\otimes \textbf{1}}$, $\textsf{Y}_{r}\in\mathbb{R}^{\mathcal{I}\otimes \textbf{1}}$ are the component tensors from the mode row block tensors $\textsf{U}$ and $\textsf{V}$ based on (\ref{eq:inverseblock}) with $K_n=R_n$ for both tensors,
%with $K_n=R_n$  and $K_n=R_n$, respectively, 
and can be viewed as the left- and right-singular tensors corresponding to the singular value $\sigma_r$. All the singular values are arranged in descending order.

\subsection{TT-algebra}
Computing the tensor algebraic notions related to even-order paired tensors using the unfolding $\psi$ and matrix operations can be computationally demanding, especially when the size of tensors is large. Oseledets \cite{doi:10.1137/090752286} showed that the basic linear algebra, such as addition, matrix-by-vector product and norms, can be done in the TT/GTT/QTT-format, without needing to go to the full tensor representation. The results were also extended for solving system of linear equations \cite{doi:10.1137/100818893,doi:10.1137/110833142}, and computation of Moore-Penrose (MP) inverse of unfolding matrices \cite{Klus_2018}. For simplicity, we summarize the key even-order paired tensor computations in the TT/GTT-format, and all the results also hold for the QTT-format.

Suppose that an even-order paired tensor $\textsf{A}\in\mathbb{R}^{\mathcal{J}\otimes\mathcal{I}}$ is given in the  GTT-format with cores $\textsf{A}^{(n)}$ and GTT-ranks $\mathcal{R}$, the U-transpose of \textsf{A}, denoted by $\textsf{A}^\top\in\mathbb{R}^{\mathcal{I}\otimes \mathcal{J}}$, can be obtained by transposing each component, i.e.,
\begin{equation}
\textsf{A}^\top = \sum_{\textbf{r}=\textbf{1}}^\mathcal{R}(\textsf{A}^{(1)}_{r_0::r_1})^\top\circ (\textsf{A}^{(2)}_{r_1::r_2})^\top\circ \dots \circ (\textsf{A}^{(N)}_{r_{N-1}::r_N})^\top.
\end{equation}
We refer to an even-order ``square'' tensor $\textsf{A}$ as weakly symmetric if $\textsf{A}=\textsf{A}^\top$. Given two even-order paired tensors $\textsf{A},\textsf{B}\in\mathbb{R}^{\mathcal{J}\otimes \mathcal{I}}$ in the  GTT-format with cores $\textsf{A}^{(n)}, \textsf{B}^{(n)}$ and GTT-ranks $\mathcal{R},\mathcal{S}$, respectively, the element-wise TT-summation is given by
\begin{equation}
\textsf{A}+\textsf{B} = \sum_{\textbf{t}=\textbf{1}}^{\mathcal{T}}\textsf{S}^{(1)}_{t_0::t_1}\circ \textsf{S}^{(2)}_{t_1::t_2}\circ \dots\circ\textsf{S}^{(N)}_{t_{N-1}::t_N}\in\mathbb{R}^{\mathcal{J}\otimes\mathcal{I}},
\end{equation}
where, $\textsf{S}^{(n)}_{t_{n-1}::t_n}$  are equal to $\textsf{A}_{r_{n-1}::r_n}$ for $t_{n-1|n}=1,2,\dots,R_{n-1|n}$, are equal to $\textsf{B}_{s_{n-1}::s_n}$ for $t_{n-1|n}=R_{n-1|n}+1,R_{n-1|n}+2,\dots, R_{n-1|n}+S_{n-1|n}$, and are equal to zero matrices otherwise, 
%\begin{equation*}
%\textsf{S}^{(n)}_{t_{n-1}::t_n} =\begin{cases} \textsf{A}_{r_{n-1}::r_n},&t_{n-1}=1,2,\dots,R_{n-1},t_{n}=1,2,\dots,R_{n} \\\textsf{B}_{s_{n-1}::s_n}, &t_{n-1}=R_{n-1}+1,R_{n-1}+2,\dots, R_{n-1}+S_{n-1}, \\ &t_{n}=R_{n}+1,R_{n}+2,\dots, R_{n}+S_{n}\\   \textbf{O},&\text{ otherwise}  \end{cases}
%\end{equation*}
with $T_n=R_n+S_n$ for $n=1,2,\dots,N-1$ and $T_0=T_N=1$. Given two even-order paired tensors $\textsf{A}\in\mathbb{R}^{\mathcal{J}\otimes \mathcal{K}}$ and $\textsf{B}\in\mathbb{R}^{\mathcal{K}\otimes \mathcal{I}}$ in the  GTT-format with  cores $\textsf{A}^{(n)}, \textsf{B}^{(n)}$ and GTT-ranks $\mathcal{R},\mathcal{S}$, respectively, the TT-Einstein product is given by
\begin{equation}\label{eq:8}
\textsf{A}*\textsf{B} = \sum_{\textbf{t}=\textbf{1}}^{\mathcal{T}}\textsf{E}^{(1)}_{t_0::t_1}\circ \textsf{E}^{(2)}_{t_1::t_2}\circ \dots\circ\textsf{E}^{(N)}_{t_{N-1}::t_N}\in\mathbb{R}^{\mathcal{J}\otimes\mathcal{I}},
\end{equation}
where, $\textsf{E}^{(n)}_{t_{n-1}::t_n} = \textsf{A}^{(n)}_{r_{n-1}::r_n}\textsf{B}^{(n)}_{s_{n-1}::s_n}\in\mathbb{R}^{J_n\times I_n}$, and $t_n = r_n+(s_n-1)R_n$ with $T_n=R_nS_n$. The computational and memory complexities of the TT-Einstein product (\ref{eq:8}) are estimated as $\mathscr{O}(NJ^3R^4)$ and $\mathscr{O}(NJ^2R^4)$, respectively assuming $J_n=I_n=K_n\sim J$ and $\mathcal{R}=\mathcal{S}\sim R$ where $R$ can be viewed as the effective rank of GTTD defined in \cite{tttoolbox}. Furthermore, the Density Matrix Renormalization Group (DMRG) based algorithms  proposed in \cite{doi:10.1137/110833142} even enable one to solve the multilinear systems
\begin{equation}
\textsf{A}*\textsf{X} = \textsf{B},
\end{equation}
where, $\textsf{A}\in \mathbb{R}^{\mathcal{J}\otimes \mathcal{J}}$ in the  GTT-format and $\textsf{B}\in\mathbb{R}^{\mathcal{J}}$ in the TT-format. Readers may refer to \cite{doi:10.1137/110833142}  for more details.

If the GTT-ranks of even-order paired tensors are low, all the above computations can be achieved efficiently with low memory costs. Although the TT-summation and the TT-Einstein product between two even-order paired tensors  may result in a new train with GTT-ranks larger than the optimal ones, one can apply TT-rounding (see Algorithm 2 in \cite{doi:10.1137/090752286}) to resolve the problem with computational complexity $\mathscr{O}(NJR^3)$. In the following, we present some variations for computing mode row/column block tensors, and ETSVD for even-order paired tensors in the GTT-format. Again, the results also hold for the QTT-format if one first quantizes the tensor trains.

\subsubsection{Block TT-format}
The block TT-format facilitates numerical computations for large scale problems \cite{DOLGOV20141207,Lebedeva2011}. Given two even-order paired tensors \textsf{A} and \textsf{B} of the same size in the  GTT-format, we can construct the $n$-mode row block tensor without converting \textsf{A} and \textsf{B} to the full representation by filling zeros into the $n$-th cores. The main steps are summarized in Algorithm \ref{alg:2.2}.

\begin{algorithm}[h]
\begin{algorithmic}[1]
\STATE{Given $\textsf{A},\textsf{B}\in\mathbb{R}^{\mathcal{J}\otimes \mathcal{I}}$ in the  GTT-format with cores $\textsf{A}^{(n)}, \textsf{B}^{(n)}$ and GTT-ranks $\mathcal{R},\mathcal{S}$ respectively, and an integer $1\leq n\leq N$}\\
\STATE{Let $\textbf{A}^{(n)} = \texttt{reshape}(\textsf{A}^{(n)}, R_{n-1}J_nI_n, R_{n})$, and $\textbf{B}^{(n)} = \texttt{reshape}(\textsf{B}^{(n)}, S_{n-1}J_nI_n, S_{n})$}\\
\STATE{Set $\tilde{\textbf{A}}^{(n)} = \begin{bmatrix} \textbf{A}^{(n)} \\ \textbf{O}_{\textsf{A}}\end{bmatrix}$, and $\tilde{\textbf{B}}^{(n)} = \begin{bmatrix}  \textbf{O}_{\textsf{B}}\\\textbf{B}^{(n)}\end{bmatrix}$ where $\textbf{O}_{\textsf{A}},\textbf{O}_{\textsf{B}}$ are zeros matrices of the same size with $\textbf{A}^{(n)},\textbf{B}^{(n)}$, respectively}\\
\STATE{Let $\textsf{A}^{(n)} = \texttt{reshape}(\tilde{\textbf{A}}^{(n)}, R_{n-1}, J_n, 2I_n, R_{n})$, and $\textsf{B}^{(n)} = \texttt{reshape}(\tilde{\textbf{B}}^{(n)}, S_{n-1}, J_n, 2I_n, S_{n})$}\\
\STATE{Compute the TT-summation $\textsf{A} + \textsf{B} =\begin{vmatrix} \textsf{A} & \textsf{B}\end{vmatrix}_n$}
\RETURN $\begin{vmatrix} \textsf{A} & \textsf{B}\end{vmatrix}_n$ in the  GTT-format.
\end{algorithmic}
\label{alg:2.2}
\caption{$n$-mode row block tensors}
\end{algorithm}

The algorithm can be generalized to multiple blocks but would be more expensive in both computation and memory for a large number of blocks compared to the full format block tensor construction. This is because TT-summation would keep increasing the GTT-ranks requiring more memory space to store the cores. Even though one can apply TT-rounding during each TT-summation, the total computational cost of the algorithm is still very high for a large number of blocks. In addition, different ways of blocking may return different computational times and memory storages depending on the structure of GTTD. For example, the TT-toolbox function \texttt{horzcat} introduces extra modes to build block tensors for the TT-matrix class \cite{tttoolbox}. Therefore, careful choice of blocking algorithms can accelerate computations and save memory. On the other hand, the inverse operation of $n$-mode row block tensor also can be achieved in the GTT-format, i.e., the blocks of an $n$-mode row block tensor can be obtained by blocking the unfolding matrix of the $n$-th core (the detailed algorithm is omitted here). Algorithm \ref{alg:2.2} can be extended to mode row/column block tensors given any factorization, and one can even choose the best factorization such that the mode row/column block tensor has the smallest GTT-ranks to achieve lowest memory requirements.

\subsubsection{TSVD}
Before discussing TSVD, we first introduce the notion of non-paired TT-format (NPTT-format). Given an even-order paired tensor $\textsf{A}\in\mathbb{R}^{\mathcal{J}\otimes \mathcal{I}}$ in the  GTT-format, its NPTT-format, naturally containing the information of its unfolding rank \cite{Chen_2019}, is defined to be the TTD of the permuted tensor $\textsf{A}_{\text{np}}\in\mathbb{R}^{\mathcal{J}\times \mathcal{I}}$. In the full representation,
\begin{equation*}
\textsf{A}_{\text{np}} = \texttt{permute}(\textsf{A}, [1,3,\dots,2N-1,2,4,\dots,2N]),
\end{equation*}
where, \texttt{permute} is the MATLAB permutation function. Significantly, the NPTTD of $\textsf{A}$ can be constructed by manipulating the cores $\textsf{A}^{(n)}$ without converting back to the full format \cite{tttoolbox}. Detailed algorithms of this conversion are given in Appendix \ref{appendix}, in which Algorithm \ref{alg:a1} transforms the  GTTD of an even-order paired tensor to the TTD, and Algorithm \ref{alg:a2} permutes the  TTD to the NPTTD. Note that both algorithms allow singular value truncations during the matrix SVD, which can remove redundancies and reduce the TT-ranks.

Klus et al. \cite{Klus_2018} exploited TTD to efficiently calculate the MP inverse of the matrix obtained from any chosen unfolding of a given tensor. By exploiting the relationship between the ETSVD and the unfolding matrix MP inverse, we adapt the framework of Klus et al. for the computation of the ETSVD of an even-order paired tensor, see Algorithm \ref{alg:2.1}. In step 3, we can introduce a truncation threshold to obtain a low unfolding rank approximation of \textsf{A}, i.e.,
$$
\textsf{A}_{K}=\sum_{r=1}^{K} \sigma_{r} \textsf{X}_{r}* \textsf{Y}_{r}^{^\top}.
$$
Appropriate truncations in the NPTT-conversion will return a good approximation of \textsf{A}. If such errors are negligible,  $\textsf{A}_{K}$ can be regarded as optimal. After obtaining the left- and right-singular tensors, we  can  choose a factorization of $R$, such that the total memory cost of the GTTD is the smallest possible in order to construct the mode row block tensors $\textsf{U}$ and $\textsf{V}$ in (\ref{eq33}). For computational convenience, we also prefer to set $R_N=R$ (or $R_1=R$) in the factorizations.

\begin{algorithm}[h]
\caption{Economy-size TSVD}
\label{alg:2.1}
\begin{algorithmic}[1]
\STATE{Given $\textsf{A}\in\mathbb{R}^{\mathcal{J}\otimes \mathcal{I}}$ in the  GTT-format with cores $\textsf{A}^{(n)}$ and GTT-ranks $\mathcal{R}^g$}\\
\STATE{Convert the GTTD of $\textsf{A}$ to its NPTTD with cores \\$\textsf{X}^{(n)}$ and TT-ranks $\mathcal{R}$ using Algorithm \ref{alg:a1} and \ref{alg:a2}}\\
\STATE{Compute the economy-size matrix SVD of $\bar{\textbf{X}}^{(N)}$, i.e.,  $\bar{\textbf{X}}^{(N)}  = \textbf{U}\boldsymbol{\Sigma}\textbf{V}^\top$ for $R = \text{rank}(\bar{\textbf{X}}^{(N)})$, and then let $\{\sigma_r\}_{r=1}^R = \texttt{diag}(\boldsymbol{\Sigma})$}
\STATE{Set $\textsf{X}^{(N)} = \texttt{reshape}(\textbf{U}, R_{N-1}, J_N, R)$ and $\textsf{X}^{(N+1)} = \texttt{reshape}(\textbf{V}^\top\bar{\textbf{X}}^{(N+1)}, R, I_1,R_{N+1})$}
\STATE{The left- and right-singular tensors of $\textsf{A}$ are given by
\begin{equation*}
\begin{split}
\textsf{U}_r &= \sum_{\textbf{r}^{-} = \textbf{1}}^{\mathcal{R}^{-}} \textsf{X}^{(1)}_{r_0:r_1}\circ\textsf{X}^{(2)}_{r_1:r_2}\circ \dots \circ \textsf{X}^{(N)}_{r_{N-1}:r},\\
\textsf{V}_r &= \sum_{\textbf{r}^{+} = \textbf{1}}^{\mathcal{R}^{+}} \textsf{X}^{(N+1)}_{r:r_{N+1}}\circ\textsf{X}^{(N+2)}_{r_{N+1}:r_{N+2}}\circ \dots \circ \textsf{X}^{(2N)}_{r_{2N-1}:r_{2N}},
\end{split}
\end{equation*}
where, $\textbf{r}^{-} = \{r_0,r_1,\dots,r_{N-1}\}$ and $\textbf{r}^{+} = \{r_{N+1},r_{N+2},\dots,r_{2N}\}$}
\RETURN Left- and right-singular tensors $\textsf{U}_r$ and $\textsf{V}_r$ in\\ the TT-format with singular values $\{\sigma_r\}_{r=1}^R$ of $\textsf{A}$.\\
\end{algorithmic}
\end{algorithm}

The algorithm can be used to compute non-negative U-eigenvalues and U-eigentensors (see details in \cite{Chen_2019}) for U-positive semidefinite weakly symmetric tensors. In this case, all the singular values are U-eigenvalues, and the left- and right-singular tensors are equal to each other, i.e., $\textsf{X}_r = \textsf{Y}_r$, and are the U-eigentensors corresponding to the U-eigenvalues $\sigma_r$. We will use this property to compute the U-eigenvalues and U-eigentensors of gramians, see section \ref{sec:3.1.2}.

Instead of reshaping and computing the economy-size matrix SVD of $\psi(\textsf{A})\in\mathbb{R}^{|\mathcal{J}|\times |\mathcal{I}|}$ which has at least $\mathscr{O}(\min{\{|\mathcal{J}|^2|\mathcal{I}|, |\mathcal{I}|^2|\mathcal{J}|\}})$ computational complexity, Algorithm \ref{alg:2.1} only computes a series of QR decompositions and matrix SVD's of the smaller-size left- and right-unfolding matrices of the cores. Hence, the computational complexity of the algorithm highly depends on the GTT-ranks and structure of $\textsf{A}$. Assume that  $J_n=I_n\sim J$ and the GTT-rank $\mathcal{R}^g \sim R$. If $R$ remains unchanged or decreases during the NPTT-conversion with appropriate truncations, the computational and memory complexities are estimated to be at most $\mathscr{O}(N^2J^3R^3)$ and $\mathscr{O}(NJR^2)$, respectively.  Otherwise, both complexities may increase exponentially with $N$. As we will see in section \ref{sec:3} and \ref{sec:4}, dealing with large sparse MLTI/LTI systems, computing ETSVD through TTD could offer significant computational and memory benefits.

\section{MLIT system reduction/identification}\label{sec:3}
The following multilinear time invariant (MLTI) system representation was first proposed by Chen et al. \cite{doi:10.1137/1.9781611975758.18},
\begin{align}\label{eq11}
\begin{cases}
\textsf{X}_{t+1}=\textsf{A}*\textsf{X}_{t}+\textsf{B}*\textsf{U}_{t}  \\
\textsf{Y}_{t}=\textsf{C}*\textsf{X}_{t}
\end{cases},
\end{align}
where,  $\textsf{X}_{t}\in\mathbb{R}^{\mathcal{J}}$ is the latent state space tensor, $\textsf{Y}_{t}\in\mathbb{R}^{\mathcal{I}}$ is the output tensor and $\textsf{U}_t\in \mathbb{R}^{\mathcal{K}}$ is an input/control tensor. $\textsf{A}\in\mathbb{R}^{\mathcal{J}\otimes \mathcal{J}}$, $\textsf{B}\in\mathbb{R}^{\mathcal{J}\otimes \mathcal{K}}$ and $\textsf{C}\in\mathbb{R}^{\mathcal{I}\otimes \mathcal{J}}$ are even-order paired tensors. Clearly, the MLTI system (\ref{eq11}) can be transformed into an equivalent linear time invariant (LTI) system via $\psi$. The transfer function $\textsf{G}(z)$ of (\ref{eq11}) is defined by
$
\textsf{G}(z) = \textsf{C}*(z\textsf{I}-\textsf{A})^{-1}*\textsf{B}
$
where $z$ is a complex variable, and the superscript $-1$ denotes the U-inverse operation \cite{Chen_2019}. Based on (\ref{eq11}), MLTI systems theoretic concepts including internal stability, reachability and observability are formulated in tensor forms. We recall several important notions from \cite{doi:10.1137/1.9781611975758.18}.

The MLTI system (\ref{eq11}) is said to be \textit{reachable} on $[t_0,t_1]$ if, given any initial condition $\textsf{X}_0$ and any final state $\textsf{X}_1$, there exists a sequence of inputs $\textsf{U}_t$ that steers the state of the system from $\textsf{X}_{t_0}=\textsf{X}_0$ to $\textsf{X}_{t_1}=\textsf{X}_1$. The pair $(\textsf{A},\textsf{B})$ is reachable on $[t_0,t_1]$ if and only if the \textit{reachability gramian}
\begin{equation}
\textsf{W}_r(t_0,t_1)=\sum_{t=t_0}^{t_1-1}\textsf{A}^{t_1-t-1}*\textsf{B}*\textsf{B}^{\top}*(\textsf{A}^{\top})^{t_1-t-1},
\end{equation}
which is a weakly symmetric even-order square tensor, is U-positive definite. Here $\textsf{A}^{t}=\textsf{A}*\textsf{A}*\stackrel{t}{\cdots}*\textsf{A}$. The infinite horizon reachability gramian can be computed from the tensor Lyapunov equation defined by
\begin{equation}
\textsf{W}_r - \textsf{A}*\textsf{W}_r*\textsf{A}^\top = \textsf{B}*\textsf{B}^\top.\label{eq:3.6}
\end{equation}
The results of observability can be simply obtained by the duality principle.

Chen et al. \cite{Chen_2019} also propose another equivalent MLTI representation with fewer parameters by using  GTTD for efficient representation and computations, which is given by (we omit two colons in the cores)
\begin{equation}\label{eq45}
\begin{cases}
\textsf{X}_{t+1} &= \displaystyle\sum_{\textbf{r}=\textbf{1}}^{\mathcal{R}_1} \textsf{X}_t\times \{\textsf{A}_{r_0r_1}^{(1)},\dots,\textsf{A}_{r_{N-1}r_N}^{(N)}\}\\&\displaystyle+\sum_{\textbf{r}=\textbf{1}}^{\mathcal{R}_2} \textsf{U}_t\times \{\textsf{B}_{r_0r_1}^{(1)},\dots,\textsf{B}_{r_{N-1r_N}}^{(N)}\}\\
\textsf{Y}_t  &= \displaystyle\sum_{\textbf{r}=\textbf{1}}^{\mathcal{R}_3} \textsf{X}_t\times \{\textsf{C}_{r_0r_1}^{(1)},\dots,\textsf{C}_{r_{N-1} r_N}^{(N)}\}.
\end{cases}
\end{equation}
Moreover, one may apply QTTD to further reduce the number of parameters and complexity of MLTI systems if the QTT-ranks are small. We will see that QTTD is significantly advantageous in both computation and memory when dealing with large sparse systems/datasets.

In many applications, such as high-dimensional partial differential equation problems, the states are high-dimensional tensors, while the number of inputs and outputs usually is much smaller than the number of states. It is thus desirable to approximate the large scale MLTI system with a lower-dimensional MLTI system. In this section, we propose three extensions of MLTI systems model reduction and identification methods - Higher-Order Balanced Truncation (HOBT), Higher-Order Balanced Proper Orthogonal Decomposition (HOBPOD) and Higher-Order Eigensystem Realization Algorithm (HOERA). Higher-Order Empirical Gramians (HOEG) will also be discussed. As mentioned, these model reduction and identification extensions also work for LTI systems if one first tensorizes the systems appropriately. 

\subsection{Higher-order balanced truncation}\label{sec:HOBT}
In LTI systems,  balanced truncation (BT), introduced by Moore \cite{1102568}, is one of the most effective model reduction methods for stable linear input-output systems. BT is able to preserve system properties such as stability and passivity \cite{1102945}, but the exact balancing is expensive to implement for large systems requiring $\mathscr{O}(n^3)$ computational complexity and $\mathscr{O}(n^2)$ storage consumption \cite{doi:10.1080/00207170410001713448}. We generalize BT to MLTI systems with fast computation and low storage consumption by exploiting the structure of the MLTI system (\ref{eq45}). Like BT, the essence of HOBT is to find a higher-order transformation $\textsf{P}\in\mathbb{R}^{\mathcal{J}\otimes \mathcal{J}}$ such that
\begin{equation}
\textsf{P}^{-1}*\textsf{W}_r*(\textsf{P}^{-1})^\top = \textsf{P}^\top*\textsf{W}_o*\textsf{P}=\textsf{S},
\end{equation}
where, $\textsf{S}\in\mathbb{R}^{\mathcal{J}\otimes \mathcal{J}}$ is the (unique up to permutation) U-diagonal tensor containing the Hankel singular values of the system which are independent of the transformations based on the unfolding properties. We  are first  required to solve the tensor Lyapunov equations (\ref{eq:3.6}) to obtain the reachability and observability gramians.

\subsubsection{Solving Lyapunov equations}
Solving Lyapunov equations by transforming them into large scale linear systems can be demanding both computationally and in terms of memory use. Many iterative methods for solving large scale Lyapunov equations, such as  the Smith method and alternating direction implicit (ADI) iteration methods, are discussed in \cite{doi:10.1137/0116017, WACHSPRESS198887}. Nip et al. \cite{6760167} proposed a novel approach by representing the large scale linear systems in the QTT-format, and solving the systems by the highly efficient DMRG based algorithms. We  adapt here the Nip et al. approach to solving the algebraic tensor Lyapunov equation
\begin{equation}\label{eq:18}
\textsf{X} - \textsf{A}*\textsf{X}*\textsf{A}^\top = \textsf{B},
\end{equation}
via the DMRG based solvers where $\textsf{A}, \textsf{B}\in\mathbb{R}^{\mathcal{J}\otimes \mathcal{J}}$ in the  GTT/QTT-format.

The key idea is to convert (\ref{eq:18}) into the following multilinear system
\begin{equation}
(\textsf{I} - \textsf{A}\circ\textsf{A})*\textsf{X}_{\text{np}} = \textsf{B}_{\text{np}},
\end{equation}
where, the outer product between two GTTD can be readily obtained by simply connecting the two trains at the last and first cores, and $\textsf{B}_{\text{np}}$ is converted from $\textsf{B}$ in the TT-format. The GTT-ranks of $\textsf{A}\circ\textsf{A}$ are two copies of the GTT-ranks of \textsf{A}, and if \textsf{B} is sparse, the NPTT-conversion is fast. One of the advantages of this approach is that we can obtain the NPTT-format of the solution, i.e., $\textsf{X}_{\text{np}}$, which can skip the NPTT-conversion in ETSVD for finding the Hankel tensor, see next subsection. Of course, the exact solution $\textsf{X}\in\mathbb{R}^{\mathcal{J}\otimes \mathcal{J}}$ can be recovered from $\textsf{X}_{\text{np}}$ as desired in the GTT-format.  Algorithm \ref{alg:3.1} applies the DMRG based linear TT-solver \texttt{amen\_solver2} \cite{tttoolbox} to solve multilinear systems, which has proven highly efficient for low GTT/QTT-ranks inputs. However,  there is no rigorous theoretical convergence analysis for DMRG based algorithms \cite{doi:10.1137/110833142}, so we will rely on numerical simulations to compare its performance.

\begin{algorithm}[h]
\caption{Solving tensor Lyapunov equation}
\label{alg:3.1}
\begin{algorithmic}[1]
\STATE{Given $\textsf{A}, \textsf{B}\in\mathbb{R}^{\mathcal{J}\otimes \mathcal{J}}$ in the  GTT-format (or \\ QTT-format) from (\ref{eq:18})}\\
\STATE{Convert \textsf{B} into its NPTT-format $\textsf{B}_{\text{np}}$ using Algorithm \\ \ref{alg:a1} and \ref{alg:a2}}, see Appendix \ref{appendix}\\
\STATE{ Set $\textsf{L} = \textsf{I} - \textsf{A}\circ \textsf{A}\in\mathbb{R}^{(\mathcal{J}\otimes \mathcal{J})\times (\mathcal{J}\otimes \mathcal{J})}$ where $\textsf{I}$ is the U-identity tensor}\\
\STATE{Apply \texttt{amen\_solver2} to solve the multilinear \\ system $\textsf{L}*\textsf{X}_{\text{np}} = \textsf{B}_{\text{np}}$}\\
\RETURN $\textsf{X}_{\text{np}}$ in the TT-format.
\end{algorithmic}
\end{algorithm}

\subsubsection{Computing Hankel singular values}\label{sec:3.1.2}
In LTI systems, after obtaining the reachability and observability gramians, computing the Hankel matrix requires computing the Schur decompositions of the gramians, which results in $\mathscr{O}(n^3)$ arithmetic operations \cite{10.1007/3-540-27909-1}. Many low-rank methods such as Krylov subspace and ADI iteration have been proposed in \cite{ HU1992283, doi:10.1137/S0895479801384937} to reduce the computational costs. In the following, we present a fast and accurate algorithm to compute the Hankel singular values using ETSVD.

The reachability and observability gramians are U-positive semidefinite weakly symmetric tensors, so computing the ETSVD will return the U-eigenvalues and U-eigentensors of the gramians. Suppose that we obtain $(\textsf{W}_r)_{\text{np}}$ and $(\textsf{W}_o)_{\text{np}}$ from Algorithm \ref{alg:3.1}. After left- and right-orthonormalizations, we can obtain the Cholesky-like factors of the two gramians, i.e.,
\begin{equation}\label{eq3.10}
\textsf{W}_r \approx \textsf{Z}_r*\textsf{Z}_r^\top \text{ and } \textsf{W}_o \approx \textsf{Z}_o*\textsf{Z}_o^\top,
\end{equation}
where, $\textsf{Z}_r\in\mathbb{R}^{\mathcal{J}\otimes\mathcal{C}}$ and $\textsf{Z}_o\in\mathbb{R}^{\mathcal{J}\otimes\mathcal{O}}$ are the mode row block tensors in the  GTT-format for $|\mathcal{C}|\leq |\mathcal{J}|$ and $|\mathcal{O}|\leq |\mathcal{J}|$ consisting of the U-eigentensors multiplied by the square root of the corresponding U-eigenvalues, respectively. The computational complexity is estimated as $\mathscr{O}(NJR^3)$ assuming $J_n\sim J$ and the TT-ranks of $(\textsf{W}_r)_{\text{np}},(\textsf{W}_o)_{\text{np}}\sim R$. In fact, if the original MLTI system possesses low GTT/QTT-ranks, $|\mathcal{C}|,|\mathcal{O}|\ll |\mathcal{J}|$. It is therefore more computationally convenient to define the Hankel tensor in the full format, i.e.,
\begin{equation}
\textsf{H} = \zeta(\textsf{Z}_o^\top)\zeta(\textsf{Z}_r)\in\mathbb{R}^{\mathcal{O}\otimes \mathcal{C}}.
\end{equation}
Then the Hankel singular values can be found more efficiently by reshaping and economy-size matrix SVD instead of using Algorithm \ref{alg:2.1}. After calculating the Hankel singular values, HOBT can be achieved as described in Algorithm \ref{alg:3.2}. In step 4, the left- and right-orthonormalization algorithms can be found in \cite{Klus_2018}. In step 5 and 7, one can choose $C_N$, $O_N$ and $S_N$ equal to the corresponding unfolding ranks, respectively in the factorizations to return the equivalent LTI representation $\textbf{A}_r\in\mathbb{R}^{|\mathcal{S}|\times |\mathcal{S}|}$, $\textbf{B}_r\in\mathbb{R}^{|\mathcal{S}\times |\mathcal{K}|}$ and $\textbf{C}_r\in\mathbb{R}^{|\mathcal{I}|\times|\mathcal{S}|}$. 

\begin{algorithm}[h]
\caption{Higher-order balanced truncation}
\label{alg:3.2}
\begin{algorithmic}[1]
\STATE{Given the MLTI system (\ref{eq11}) with \textsf{A}, \textsf{B} and \textsf{C} in the\\ full format}\\
\STATE{Convert the system to (\ref{eq45}) using  GTTD (or QTTD)}
\STATE{Compute the reachability and observability gramians $(\textsf{W}_r)_{\text{np}}$ and $(\textsf{W}_o)_{\text{np}}$ by Algorithm \ref{alg:3.1} from the tensor Lyapunov equations}\\
\STATE{Left-orthonormalize the first $N-1$ cores and right-orthonormalize the last $N$ cores of $(\textsf{W}_r)_{\text{np}}$ and $(\textsf{W}_o)_{\text{np}}$, respectively}\\
\STATE{ Apply step 3-5 from Algorithm \ref{alg:2.1} to obtain $\textsf{W}_r \approx \textsf{Z}_r*\textsf{Z}_r^\top$ and $\textsf{W}_o \approx \textsf{Z}_o*\textsf{Z}_o^\top$ }\\
\STATE{Compute the Hankel tensor $\textsf{H} = \zeta(\textsf{Z}_o)^\top* \zeta(\textsf{Z}_r)$ in \\ the full format}\\
\STATE{Compute the best  unfolding rank $S$ approximation \\ of $\textsf{H}$ using reshaping and economy-size matrix SVD, i.e., $\textsf{H} \approx \textsf{U}*\textsf{S}*\textsf{V}^\top$ with $\textsf{U}\in\mathbb{R}^{\mathcal{O}\otimes \mathcal{S}}$, $\textsf{V}\in\mathbb{R}^{\mathcal{C}\otimes \mathcal{S}}$, $\textsf{S}\in\mathbb{R}^{\mathcal{S}\otimes \mathcal{S}}$ where $\mathcal{S} = \{S_1,S_2, \dots, S_N\}$}\\
\STATE{Calculate $\textsf{P} = \zeta(\textsf{Z}_r)*\textsf{V}*\textsf{S}^{-\frac{1}{2}}\in\mathbb{R}^{\mathcal{J}\otimes \mathcal{S}}$ and $\textsf{Q} = \zeta(\textsf{Z}_o)*\textsf{U}*\textsf{S}^{-\frac{1}{2}}\in\mathbb{R}^{\mathcal{J}\otimes \mathcal{S}}$ in the full format}\\
\STATE{The reduced model is given by $\textsf{A}_r = \textsf{Q}^\top*\textsf{A}*\textsf{P}$, $\textsf{B}_r = \textsf{Q}^\top*\textsf{B}$ and $\textsf{C}_r = \textsf{C}*\textsf{P}$}\\
\RETURN  Reduced MLTI system $\textsf{A}_r\in\mathbb{R}^{\mathcal{S}\otimes \mathcal{S}}$, \\$\textsf{B}_r\in\mathbb{R}^{\mathcal{S}\otimes \mathcal{K}}$ and $\textsf{C}_r\in\mathbb{R}^{\mathcal{I}\otimes \mathcal{S}}$.\\
\end{algorithmic}
\end{algorithm}

\subsubsection{Error bounds}
The regular BT provides an a priori error bound for the reduced system based on the Hankel singular values \cite{4047847}. However, this error bound does not hold  exactly for HOBT because several truncation errors occur during the GTTD/QTTD, the DMRG solver and the ETSVD. We nevertheless can estimate an error bound for HOBT based on the exact balancing. Suppose \textsf{G} and $\textsf{G}_r$ are the transfer functions of the full MLTI and reduced MLTI systems, respectively. Based on the triangular inequality, we have
\begin{equation}\label{eq:3.11}
\|\textsf{G} - \textsf{G}_{r}\|_\infty = \|\psi(\textsf{G}) - \psi(\textsf{G}_{r})\|_\infty \leq 2\sum_{r=S+1}^{|\mathcal{J}|}\sigma_r + \epsilon,
\end{equation}
where, $\sigma_r$ are the exact Hankel singular values, $S$ is the ETSVD truncation point, and $\epsilon$ is the $\mathcal{H}$-$\infty$ norm error between the two reduced systems from the exact balancing and HOBT. Knowing that computing the exact error between \textsf{G} and $\textsf{G}_{r}$ can be computationally intense for large scale systems, (\ref{eq:3.11}) can be used to obtain an estimate of the performance of HOBT more effectively. 

HOBT depends on the solutions of the tensor Lyapunov equations, but the DMRG based solvers are very sensitive to the GTT/QTT-ranks and the structure of MLTI systems. Therefore, more efficient methods for approximating gramians are required.

\subsection{Higher-order empirical gramians}\label{sec:HOEG}
Instead of computing the gramians by solving the tensor Lyapunov equation (\ref{eq:18}) using Algorithm \ref{alg:3.1}, one may directly compute them from data or numerical simulations \cite{doi:10.1142/S0218127405012429}. This was the original approach used in Lall et al. \cite{LALL19992598, doi:10.1002/rnc.657} to extend balanced truncation to nonlinear systems. To obtain the higher-order empirical reachability gramain of a MLTI system with input size $\mathcal{K}$, one runs the ``higher-order impulse response'' simulations of the primal MLTI system (\ref{eq11}).  Writing $\textsf{B}=\begin{vmatrix} \textsf{B}_1 & \textsf{B}_2 & \dots & \textsf{B}_{|\mathcal{K}|}\end{vmatrix}$ based on the inverse operation of mode row block tensor construction (\ref{eq:inverseblock}), one constructs the state responses over $t=0,1,\dots,T$ as follows:
\begin{equation}\label{eq:4.4}
\textsf{X}^{(i)} =\begin{vmatrix}
\textsf{B}_i & \textsf{A}*\textsf{B}_i & \dots & \textsf{A}^{T}*\textsf{B}_i\end{vmatrix}.
\end{equation}
Consider $T+1=T_1T_2\dots T_N$ snapshots of states for $i = 1,2,\dots,|\mathcal{K}|$, and arrange the snapshots in the following form:
\begin{equation}\label{eq:4.5}
\textsf{X} = \begin{vmatrix}
\textsf{X}^{(1)} & \textsf{X}^{(2)} & \dots & \textsf{X}^{(|\mathcal{K}|)}\end{vmatrix}\in\mathbb{R}^{\mathcal{J}\otimes\mathcal{K}\mathcal{T}}.
\end{equation}
The higher-order empirical reachability gramian is given by
\begin{equation}\label{eq26}
\textsf{W}_r\approx \textsf{X}*\textsf{X}^\top.
\end{equation}

This gramian is consistent with the definition in LTI systems. In other words, the unfolding $\psi(\textsf{W}_r)$ is the empirical reachability gramian of the LTI system $\big(\psi(\textsf{A}), \psi(\textsf{B}), \psi(\textsf{C})\big)$. According to Proposition 2.2 in \cite{doi:10.1137/1.9781611975758.18}, it follows that
$
\psi(\textsf{X}^{(i)}) = \begin{bmatrix}
\textbf{b}_i & \textbf{A}\textbf{b}_i & \dots & \textbf{A}^{T}\textbf{b}_i\end{bmatrix} = \textbf{R}_i
$
where $\textbf{b}_i=\psi(\textsf{B}_i)$ and $\textbf{A} = \psi(\textsf{A})$. Then
\begin{equation*}
\begin{split}
\psi(\textsf{X}) &=  \begin{bmatrix}
\psi(\textsf{X}^{(1)}) &\psi(\textsf{X}^{(2)}) & \dots & \psi(\textsf{X}^{(|\mathcal{K}|)})\end{bmatrix}\textbf{P}\\&=\begin{bmatrix} \textbf{R}_1 & \textbf{R}_2 & \dots & \textbf{R}_{|\mathcal{K}|}\end{bmatrix}\textbf{P}=\textbf{X}\textbf{P},
\end{split}
\end{equation*}
for a column permutation matrix $\textbf{P}$. Hence, $\psi(\textsf{W}_c)\approx\psi(\textsf{X})\psi(\textsf{X})^\top = \textbf{X}\textbf{P}\textbf{P}^\top\textbf{X}^\top=\textbf{X}\textbf{X}^\top$. On the other hand, the procedure proceeds similarly for constructing the output snapshot tensor by collecting output snapshots $\textsf{Y}\in\mathbb{R}^{\mathcal{J}\otimes \mathcal{I}\mathcal{L}}$ from the simulations of the adjoint MLTI system
\begin{equation}\label{eq:4.11}
\tilde{\textsf{X}}_{t+1} = \textsf{A}^\top*\tilde{\textsf{X}}_t + \textsf{C}^\top*\textsf{V}_t,
\end{equation}
over $t=0,1,\dots,L$ such that $L+1 = L_1L_2\dots L_N$.

Suppose we are only given the MLTI system (\ref{eq11}) to begin with. After converting (\ref{eq11}) to (\ref{eq45}) by GTTD (or QTTD) with low GTT-ranks (or QTT-ranks), all the steps (\ref{eq:4.4}-\ref{eq26}) described above can be efficiently achieved using TT-Einstein product and block TT-format for small values of $T,L$. However, if the impulse state responses data is provided for large $T$ and $L$, it may be more efficient to construct \textsf{X} and \textsf{Y} using mode row block tensor in the full format.
%Based on the formulation of HOEG, we can discuss about HOBPOD.

\subsubsection{Higher-order balanced POD}\label{sec:HOBPOD}
The method of balanced proper orthogonal decomposition (BPOD) was first proposed by Rowley \cite{doi:10.1142/S0218127405012429} to deal with model reduction problems in high-dimensional input/output spaces such as in the context of fluid mechanics. Analogously, the goal of HOBPOD is to obtain an approximate HOBT that is computationally tractable for a large scale MLTI system. The method includes computing the balancing transformation from the collections of snapshots and a higher-order output projection method using ETSVD.

Suppose that the reachability and observability snapshots $\textsf{X}$ and $\textsf{Y}$ are given as in HOEG.  Then the generalized Hankel tensor is defined by
\begin{equation}
\textsf{H} = \textsf{Y}^\top*\textsf{X}\in\mathbb{R}^{\mathcal{I}\mathcal{L}\otimes \mathcal{K}\mathcal{T}}.
\end{equation}
We can then simply apply Algorithm \ref{alg:2.1} to obtain the generalized Hankel singular values, and the remaining steps proceed similarly as HOBT. In Algorithm \ref{alg:3.3}, we provide the case where the higher-order impulse state response data is given. In step 2, we use mode row block tensor to construct $\textsf{X}$ and $\textsf{Y}$, but other choices of tensor blocking can also be used (e.g., extending the modes to build block tensors). Different choices of blocking would result in different computing times and memory storage requirements in the GTTD and ETSVD, but if one prefers to use QTTD, the choice of tensor blocking approach may not result in a significant difference. In particular, one can even use block matrices to form the generalized Hankel matrix $\textbf{H}$ using $\psi$ and apply QTTD directly to $\textbf{H}$. In step 4, one can also permute \textsf{H} to $\textsf{H}_{\text{np}}$ first in the full format and then apply QTTD to $\textsf{H}_{\text{np}}$, but this approach may lose the flexibility of singular value truncations during the NPTT-conversion. In fact, it may be slower in computing the generalized Hankel singular values for similar level of accuracy. Additionally, depending on the situation, it may not be efficient to use the TT-Einstein product in step 4, 6 and 7 to compute the generalized Hankel tensor \textsf{H}, transformations \textsf{P},\textsf{Q} and the reduced system $\textsf{A}_r,\textsf{B}_r,\textsf{C}_r$. For instance, in very high-dimensional MLTI systems, the GTT/QTT-ranks could be relatively large even if they are much less than the full ones, and so the TT-Einstein product may become a computational bottleneck.

HOBPOD requires access to the MLIT system representation and its adjoint, and thus is not applicable in experimental setting. Moreover, the construction of the generalized Hankel tensor from the Einstein product between the reachability and observability snapshots $\textsf{X}$ and $\textsf{Y}$ can be computationally costly for high-dimensional MLTI systems. The eigensystem realization algorithm based approach alleviates these limitations of HOBPOD/BPOD.

\begin{algorithm}[h]
\caption{Higher-order balanced POD}
\label{alg:3.3}
\begin{algorithmic}[1]
\STATE{Given the snapshots of state responses $\textsf{X}_t\in\mathbb{R}^{\mathcal{J}}$ and $\hat{\textsf{X}}_l\in\mathbb{R}^{\mathcal{J}}$, two integer $T,L$ and the MLTI system (\ref{eq11}) \\with \textsf{A}, \textsf{B}, \textsf{C} in the full format}\\
\STATE{Construct the reachability snapshot tensor using mode row block tensor
$$
\textsf{X} = \begin{vmatrix} \textsf{X}^{(1)} & \textsf{X}^{(2)} & \dots & \textsf{X}^{(|\mathcal{K}|)}\end{vmatrix},
$$
where, $\textsf{X}^{(k)} = \begin{vmatrix}  \textsf{X}_0 & \textsf{X}_1 & \dots & \textsf{X}_T\end{vmatrix}$ with $T+1=|\mathcal{T}|$
}
\STATE{Construct the observability snapshot tensor $\textsf{Y}$ using \\ the adjoint state responses $\hat{\textsf{X}}_l$ similarly as step 2}
\STATE{Compute the generalized Hankel tensor $\textsf{H}=\textsf{Y}^\top*\textsf{X}$ \\and apply GTTD (or  QTTD) to \textsf{H}}\\
\STATE{Compute the unfolding rank $S$ approximation of $\textsf{H}$\\ using Algorithm \ref{alg:2.1}, i.e., $\textsf{H} \approx \textsf{U}*\textsf{S}*\textsf{V}^\top$ with $\textsf{U}\in\mathbb{R}^{\mathcal{I}\mathcal{L}\otimes \mathcal{S}}$, $\textsf{V}\in\mathbb{R}^{\mathcal{K}\mathcal{T}\otimes \mathcal{S}}$, $\textsf{S}\in\mathbb{R}^{\mathcal{S}\otimes \mathcal{S}}$ where $\mathcal{S} = \{S_1,S_2, \dots, S_N\}$}\\
\STATE{Follow similarly as step 8 and 9 in Algorithm \ref{alg:3.2}}\\
\RETURN  Reduced MLTI system $\textsf{A}_r\in\mathbb{R}^{\mathcal{S}\otimes \mathcal{S}}$, \\$\textsf{B}_r\in\mathbb{R}^{\mathcal{S}\otimes \mathcal{K}}$ and $\textsf{C}_r\in\mathbb{R}^{\mathcal{I}\otimes \mathcal{S}}$.\\
\end{algorithmic}
\end{algorithm}

\subsection{Higher-order ERA}\label{sec:HOERA}
The eigensystem realization algorithm (ERA) was first proposed in \cite{doi:10.2514/3.20031} as a model identification and reduction tool for LTI systems.  It was later shown \cite{Ma2011} that ERA is theoretically equivalent to  BPOD for discrete time systems but with significantly lower computational cost. Similarly, the idea of HOERA is to construct a generalized Hankel tensor using higher-order impulse response simulations or experiments without having access to the MLTI system (\ref{eq11}). 

First, we need to collect the snapshots of the impulse response $\textsf{Z}_k\in\mathbb{R}^{\mathcal{I}\otimes \mathcal{K}}$ from simulations/experiments and form the generalized Hankel tensor as follows:
\begin{equation}\label{eq:3.17}
\textsf{H}=
\begin{vmatrix}
\textsf{Z}_0 & \textsf{Z}_1 & \dots & \textsf{Z}_T\\
\textsf{Z}_1 & \textsf{Z}_2 & \dots & \textsf{Z}_{T+1}\\
\vdots & \vdots & \ddots & \vdots\\
\textsf{Z}_L & \textsf{Z}_{L+1} & \dots & \textsf{Z}_{T+L}
\end{vmatrix}\in\mathbb{R}^{\mathcal{I}\mathcal{L}\otimes \mathcal{K}\mathcal{T}},
\end{equation}
where, $\textsf{Z}_k = \textsf{C}*\textsf{A}^k*\textsf{B}$ are called Markov parameters, $T+L+2$ is the number of snapshots with $T+1=T_1T_2\dots T_N$ and $L+1 = L_1L_2\dots L_N$. $T$ and $L$ are usually chosen to be sufficiently large. By applying Algorithm \ref{alg:2.1}, we can obtain the generalized Hankel singular values, and the remaining steps are summarized in Algorithm \ref{alg:3.4}. Like HOBPOD, the QTTD of \textsf{H} is computationally less affected by different choices of blocking. We can also convert the tensors to the full format after ETSVD in computing the reduced model at step 4 in case the TT-Einstein product becomes a computational bottleneck.

The reduced MLTI system obtained from HOERA is consistent with the one obtained by applying ERA on the unfolded LTI system. In other words, the reduced system $\big(\psi(\textsf{A}_r)$, $\psi(\textsf{B}_r)$, $\psi(\textsf{C}_r)\big)$ can be achieved from ERA on the LTI system $\big(\psi(\textsf{A}), \psi(\textsf{B}), \psi(\textsf{C})\big)$. Based on Proposition 2.2 in \cite{doi:10.1137/1.9781611975758.18}, it follows that
\begin{equation*}
\psi(\textsf{H}) = \textbf{Q}
\begin{bmatrix}
\psi(\textsf{Z}_0) & \psi(\textsf{Z}_1) & \dots & \psi(\textsf{Z}_T)\\
\psi(\textsf{Z}_1) & \psi(\textsf{Z}_2) & \dots & \psi(\textsf{Z}_{T+1})\\
\vdots & \vdots & \ddots & \vdots\\
\psi(\textsf{Z}_L) & \psi(\textsf{Z}_{L+1}) & \dots & \psi(\textsf{Z}_{T+L})
\end{bmatrix}\textbf{P}=\textbf{Q}\textbf{H}\textbf{P},
\end{equation*}
where, $\textbf{Q}$ is a row permutation matrix and $\textbf{P}$ is a column permutation matrix. Suppose that the matrix SVD of $\textbf{H}$ is given by $\textbf{H} = \textbf{U}\textbf{S}\textbf{V}^\top$. Then $\psi(\textsf{U}) = \textbf{Q}\textbf{U}$, $\psi(\textsf{V}) = \textbf{P}^\top\textbf{V}$ and $\psi(\textsf{S}) = \textbf{S}$. Hence,
\begin{equation*}
\begin{split}
\psi(\textsf{A}_r) &= \textbf{S}^{-\frac{1}{2}}\textbf{U}^\top\textbf{Q}^\top\textbf{Q}\textbf{H}_1\textbf{P}\textbf{P}^\top\textbf{V} \textbf{S}^{-\frac{1}{2}} = \textbf{S}^{-\frac{1}{2}}\textbf{U}^\top\textbf{H}_1\textbf{V} \textbf{S}^{-\frac{1}{2}},\\
\psi(\textsf{B}_r) & = \textbf{S}^{-\frac{1}{2}}\textbf{U}^\top\textbf{Q}^\top\textbf{Q}\text{Col}_{\text{first}}(\textbf{H}) = \textbf{S}^{-\frac{1}{2}}\textbf{U}^\top\text{Col}_{\text{first}}(\textbf{H}),\\
\psi(\textsf{C}_r)  &= \text{Row}_{\text{first}}(\textbf{H})\textbf{P}\textbf{P}^\top\textbf{V} \textbf{S}^{-\frac{1}{2}}  = \text{Row}_{\text{first}}(\textbf{H})\textbf{V} \textbf{S}^{-\frac{1}{2}},
\end{split}
\end{equation*}
where, $\psi(\textsf{H}_1) = \textbf{Q}\textbf{H}_1\textbf{P}$, and $\text{Col}_{\text{first}}(\textbf{H})$ and $\text{Row}_{\text{first}}(\textbf{H})$ represent the first block column and row of the generalized Hankel matrix $\textbf{H}$.

\begin{algorithm}[h]
\caption{Higher-order ERA}
\label{alg:3.4}
\begin{algorithmic}[1]
\STATE{Given the snapshots of the impulse response $\textsf{Z}_k\in\mathbb{R}^{\mathcal{I}\otimes \mathcal{K}}$, two integers $T,L$}
\STATE{Construct the generalized Hankel tensor $\textsf{H}$ using \\mode block tensor with $T+1=|\mathcal{T}|$ and $L+1=|\mathcal{L}|$ and apply  GTTD (or  QTTD) to $\textsf{H}$}\\
\STATE{Compute the unfolding rank $S$ approximation of $\textsf{H}$\\ using Algorithm \ref{alg:2.1}, i.e., $\textsf{H} \approx\textsf{U}*\textsf{S}*\textsf{V}^\top$}\\
\STATE{The reduced model is given by \begin{align*}&\textsf{A}_r = \zeta(\textsf{S}^{-\frac{1}{2}})*\zeta(\textsf{U}^\top)*\textsf{H}_1*\zeta(\textsf{V})*\zeta(\textsf{S}^{-\frac{1}{2}}),\\ &\textsf{B}_r = \zeta(\textsf{S}^{-\frac{1}{2}})*\zeta(\textsf{U}^\top)*\text{Col}_{\text{first}}(\textsf{H}),\\ &\textsf{C}_r= \text{Row}_{\text{first}}(\textsf{H})*\zeta(\textsf{V})*\zeta(\textsf{S}^{-\frac{1}{2}})\end{align*} in the full format where $\text{Col}_{\text{first}}(\textsf{H})$ and $\text{Row}_{\text{first}}(\textsf{H})$ represent the first block column and row of $\textsf{H}$ and
\begin{equation*}
\textsf{H}_1=
\begin{vmatrix}
\textsf{Z}_1 & \textsf{Z}_2 & \dots & \textsf{Z}_{T+1}\\
\textsf{Z}_2 & \textsf{Z}_3 & \dots & \textsf{Z}_{T+2}\\
\vdots & \vdots & \ddots & \vdots\\
\textsf{Z}_{L+1} & \textsf{Z}_{L+2} & \dots & \textsf{Z}_{T+L+1}
\end{vmatrix}\in\mathbb{R}^{\mathcal{I}\mathcal{L}\otimes \mathcal{K}\mathcal{T}}
\end{equation*}
}
\RETURN  Reduced MLTI system $\textsf{A}_r\in\mathbb{R}^{\mathcal{S}\otimes \mathcal{S}}$,\\ $\textsf{B}_r\in\mathbb{R}^{\mathcal{S}\otimes \mathcal{K}}$ and $\textsf{C}_r\in\mathbb{R}^{\mathcal{I}\otimes \mathcal{S}}$.\\
\end{algorithmic}
\end{algorithm}

Moreover, if the output $\textsf{Z}_k$ are collected from ``higher-order step response'' simulations or experiments, HOERA can still be applied. It can be shown that the difference between two consecutive outputs is given by
\begin{equation*}
\delta \textsf{Z}_k = \textsf{Z}_k - \textsf{Z}_{k-1} = \textsf{C}*\textsf{A}^{k-1}*\textsf{B},
\end{equation*}
and one can simply apply Algorithm \ref{alg:3.4} to the snapshots $\{\delta \textsf{Z}_1, \delta \textsf{Z}_2, \dots, \delta \textsf{Z}_{T+L+1}\}$ to obtain the reduced MLTI model.

\section{Numerical examples}\label{sec:4}
All the numerical examples presented were performed on a Windows 7 desktop with 16 GB RAM and a 3.3 GHz Intel Core i7 processor and were conducted in MATLAB 2018b with the TT-Toolbox 2.2 by Oseledets et al. \cite{tttoolbox}.

\subsection{Synthetic datasets}
\begin{figure}[tbhp]
\centering
\includegraphics[scale=0.3]{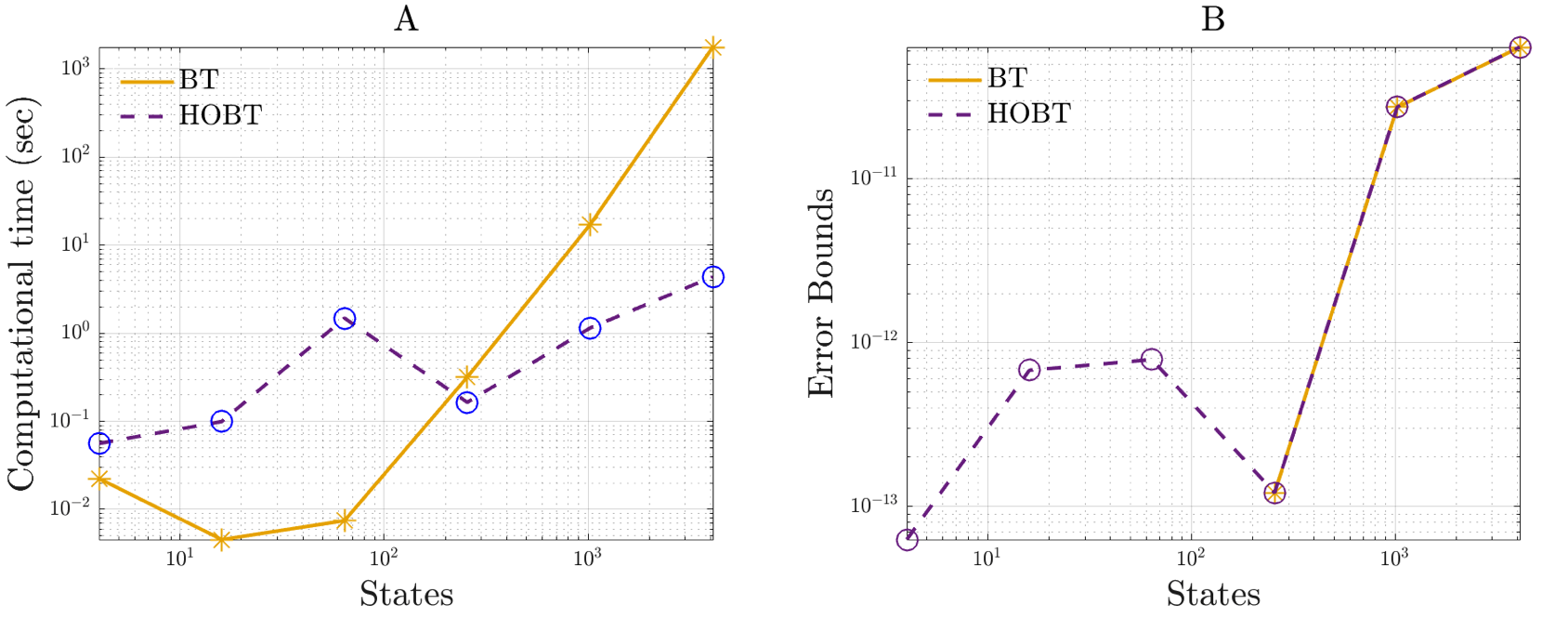}
\caption{Computational time (A) and error bound (B) comparisons between BT and HOBT. The error bounds for the first three states of BT are zeros, so they are not shown on the plot. }
\label{fig:4.5}
\end{figure}

In this example, we consider a multiple input multiple output (MIMO) MLTI system (\ref{eq45}) with low QTT-ranks random tensors in the QTT-format
$\textsf{A}\in\mathbb{R}^{\textbf{2}\otimes\textbf{2}}, \textsf{B}\in\mathbb{R}^{\textbf{2}\otimes\textbf{2}}$ and $\textsf{C}\in\mathbb{R}^{\textbf{2}\otimes\textbf{2}}$. Consequently, the number of states in the unfolded LTI corresponding to MLTI system (\ref{eq11}) is $2^N$.
%$\textsf{A}\in\mathbb{R}^{\textbf{2}\times\textbf{2}}, \textsf{B}\in\mathbb{R}^{\textbf{2}\times\textbf{2}}$ and $\textsf{C}\in\mathbb{R}^{\textbf{2}\times\textbf{2}}$ in the QTT-format.
We compare the computing times of higher-order balanced truncation (HOBT) and ordinary balanced truncation (BT) for different values of $N$ (and hence number of states) in Figure \ref{fig:4.5} (A), and the error bounds in Figure \ref{fig:4.5} (B).  As can be seen, the computing time for ordinary BT grows rapidly with the number of states $N$ but remains bounded for HOBT, while the errors in model reduction using the two approaches are similar. For the error bound, we use the sum of the residual Hankel singular values for the ordinary BT, and the bound (\ref{eq:3.11}) for HOBT.

\subsection{2D heat equations with control}
The heat equation is a partial differential equation that describes the evolution of heat distribution in a solid medium over time. The 2D heat equation on  the square $D = [-\pi,\pi]^2$ with localized point control and Dirichlet boundary conditions is given by
\begin{equation}
\begin{cases}
\frac{\partial}{\partial t} \phi(t, \textbf{x}) = c^2 \frac{\partial^2}{\partial \textbf{x}^2} \phi(t, \textbf{x}) + \delta(\textbf{x})u_t, \text{ } \textbf{x}\in D,\\
\phi(\textbf{x}, t) = 0, \text{ } \textbf{x}\in\partial D, \label{heatEqn}
\end{cases}
\end{equation}
where, $c>0$, $u_t\in\mathbb{R}$ is a one-dimensional control input and $\delta(\textbf{x})$ is the Dirac delta function centered at zeros. In this example, we apply higher-order balanced proper orthogonal decomposition (HOBPOD) and higher-order eigensystem realization algorithm (HOERA) to the discretized  heat equation (\ref{heatEqn}) to find the reduced balanced system.

We use a second-order central difference to approximate the Laplacian, first-order difference in time, and we approximate the Dirac delta as a Kronecker delta function at the nearest grid point with unit mass, leading to a MTLI system of the form (\ref{eq11}), where $\textsf{X}_t \in \mathbb{R}^{N\times N}$ is the 2D-temperature field at instance $t$, and $\textsf{A}\in \mathbb{R}^{N\times N \times N\times N}$, $\textsf{B}\in \mathbb{R}^{N\times I_1\times N\times I_2}$ and $\textsf{C}\in \mathbb{R}^{I_1\times N \times I_2 \times N}$ are system tensors. The tensor $\textsf{A}$ is the tensorization
%of matrix $\textbf{A}$
%\begin{equation}\label{eq:}
%\textbf{A}=\textbf{I}_{\textbf{N}^2}-\textbf{I}_{\textbf{N}},
%\end{equation}
of matrix $\frac{c^2 \Delta t}{h^2}\mathbf{\Delta}_{dd}\in \mathbb{R}^{\textbf{N}^2\times\textbf{N}^2}$, where $\mathbf{\Delta}_{dd}$ is the discrete Laplacian on a rectangular grid with Dirichlet boundary conditions, and $\textsf{B}$ is the tensorization of $\frac{1}{h^2}\hat{\delta}(x)$, where $\hat{\delta}(x)$ is a vector which is zero everywhere except at the entry corresponding to the grid point closest to the origin (see \cite{6760167} for details). Here, $h=\Delta x=\Delta y$ are spatial resolution of grid sizes in the $x$ and $y$ directions, respectively, $\Delta t$ is discretization in time, and we ensure the Courant-Friedrichs-Lewy (CFL) condition $\frac{ c^2 \Delta t}{h^2}<1$ is met for numerical stability.

\begin{figure}[tbhp]
\centering
\includegraphics[scale=0.34]{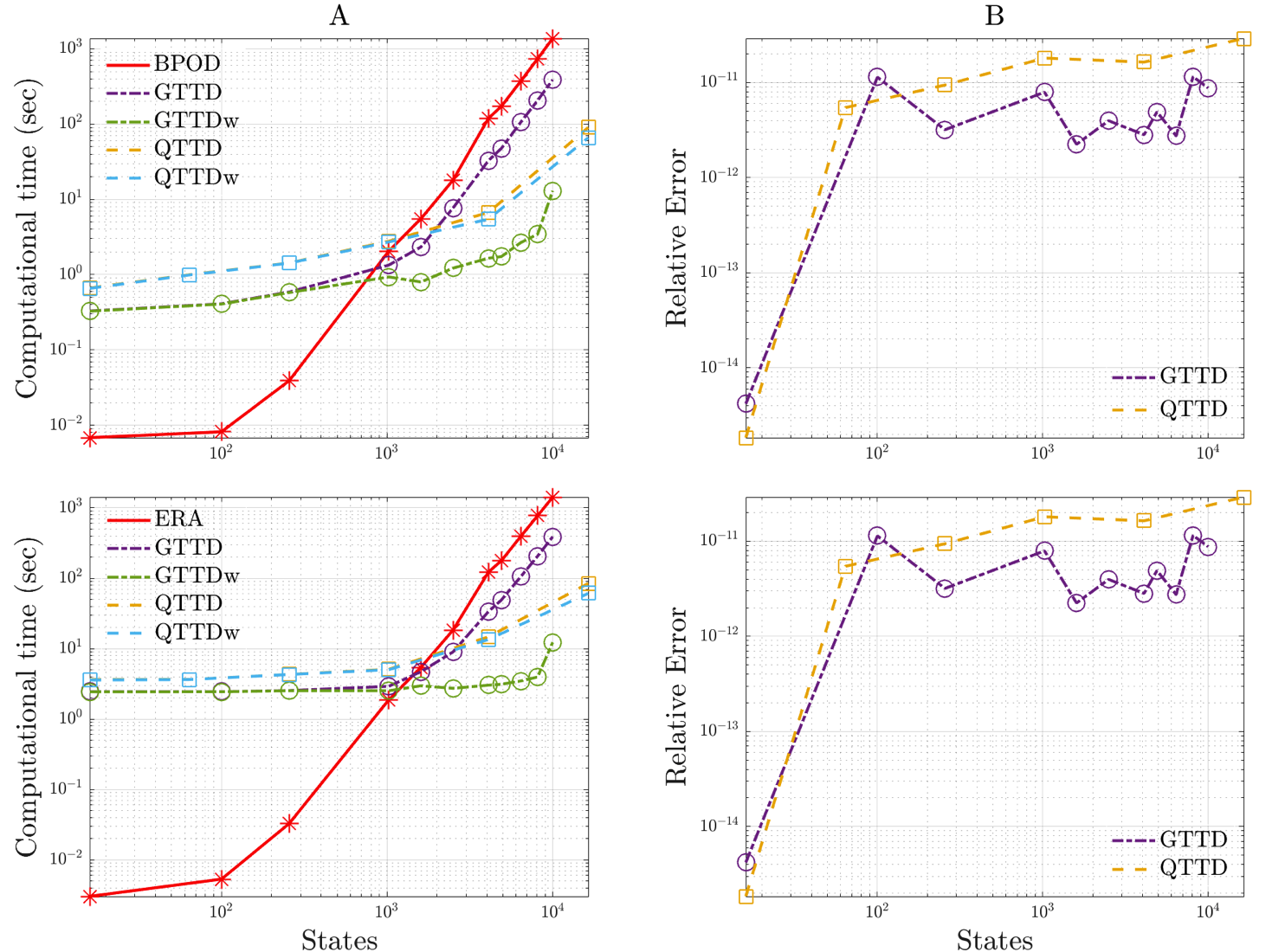}
\caption{Computational time (A) and relative error (B) comparisons between BPOD/ERA and HOBPOD/HOERA with GTTD and QTTD, in which GTTDw stands for the case where the conversion time to GTTD is omitted, and similarly for QTTDw.}
\label{fig:4.4}
\end{figure}

We assume measurement at a single discrete location, i.e., $I_1=I_2=1$,  and generate forward/adjoint input/output snapshot data over the $T+L+1$ time steps for different values of spatial discretization $h$. Let $N=2\pi/h$ be the dimensions of the two modes of the temperature tensors $\textsf{X}_t$. Figure \ref{fig:4.4} (A) shows the comparison of HOBPOD/HOERA computational times with balanced proper orthogonal decomposition/eigensystem realization algorithm (BPOD/ERA) as a function of system dimension, i.e., number of states. We consider both GTTD and QTTD approaches for HOBPOD/HOERA, and both are found to be more efficient than BPOD/ERA. Furthermore, as expected, QTTD is more efficient than GTTD if the tensor train conversion time is included. Also shown in the figure are the curves for GTTDw (and similarly QTTDw), which stands for computational time for GTTD based HOBPOD/HOERA without accounting for the time to convert the tensors to their corresponding GTT-format. This is because some special tensors/matrices like the  Laplacian can be directly and efficiently constructed in the GTT/QTT-format without requiring first constructing the matrix equivalent (e.g., see the function \texttt{tt\_qlaplace\_dd} in the MATLAB TT-Toolbox \cite{tttoolbox}). Note that while this conversion time can be significant for GTTD, for QTTD it is found to be negligible. Figure \ref{fig:4.4} (B) shows difference between transfer functions obtained via BPOD/ERA and HOBPOD/HOERA, indicating that HOBPOD/HOERA obtains the reduced model with similar accuracy to BPOD/ERA.

\subsection{Cancer cell image video dataset}
The Fas-Associated Death Domain (FADD) is an adaptor protein known for its role in cell extrinsic apoptosis (cell death), and studies have revealed that deletion of FADD is able to suppress lung cancer development \cite{cancer}. In this example, we try to capture the cell cycle dynamics of a cancer cell with and without downregulation/suppression of FADD by HOERA/ERA. The downregulation of FADD is accomplished by the intranasal inhalation of an adenovirus expressing Cre recombinase (AdCre) \cite{cancer}, which can be viewed as a step input to the cell cycle dynamics.

\begin{figure*}[tbhp]
\centering
\includegraphics[width=\textwidth]{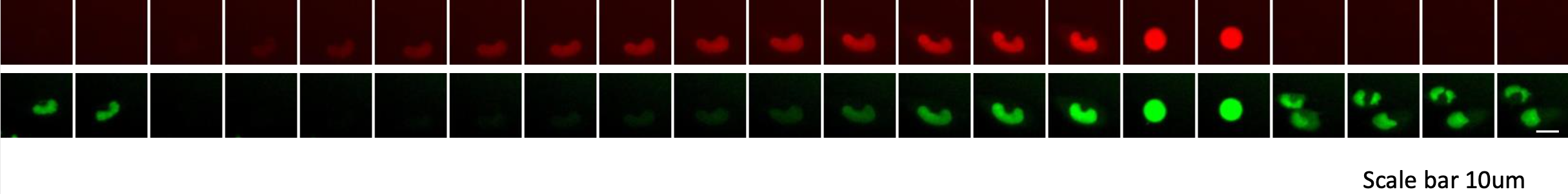}
\caption{An example of image videos for a cancer cell with/without FADD downregulation.}
\label{fig:4.6}
\end{figure*}

The dataset contains two cancer cell image videos, see Figure \ref{fig:4.6}. One is a normal cancer cell, and one is a cancer cell with deletion of FADD. Each video has 145 frames with pixels $64\times 64$ and two channels. The channels are used to indicate cell cycle state transitions. For both videos, we binarize each frame and construct the generalized Hankel matrix $\textbf{H}\in\mathbb{R}^{2^{17}\times 2^7}$ with $T=127$ and $L=15$ using $\psi$. Then we apply  QTTD to \textbf{H} and obtain a tensor train with $J_n=2$ for $n=1,2,\dots, 17$, $I_n=1$ for $n=1,2,\dots,10$ and $I_n=2$ for $n=11,12,\dots,17$. In the following study, we compare the Hankel singular values in HOERA and ERA with different truncations in the QTTD step and observe the dominant features from the first few left-singular tensors and left-singular vectors, respectively, for the two videos. Along the way, we report the number of parameters that require to store the outputs of ETSVD and economy-size matrix SVD.

\begin{figure*}[tbhp]
\centering
\includegraphics[width=\textwidth]{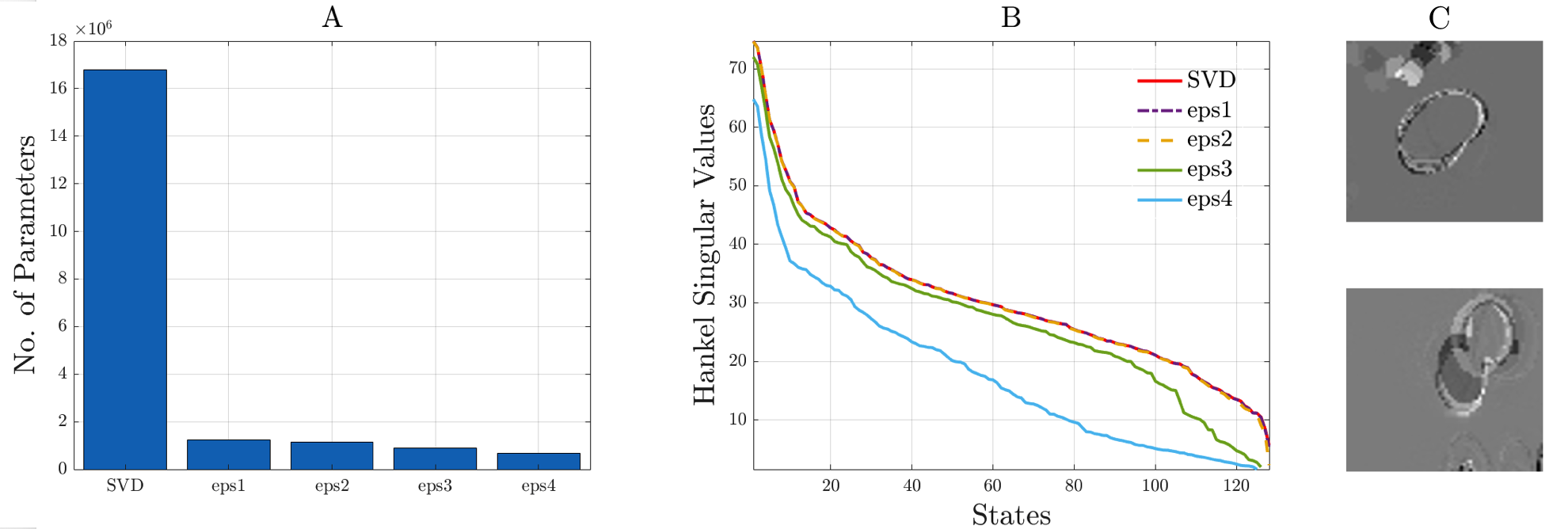}
\caption{(A) plots the number of parameters produced by economy-size matrix SVD and ETSVD with $\epsilon=0.01,0.1,0.5,1$ in computing the QTTD for FADD deletion video. (B) plots the Hankel singular values obtained from matrix SVD and ETSVD with different $\epsilon$. (C) contains two images of the first left-singular tensors at the 16 frame for FADD deletion video and normal cancer cell video, respectively, with $\epsilon = 0.1$.}
\label{fig:4.2}
\end{figure*}

The results are shown in Figure \ref{fig:4.2}. The ETSVD has a huge advantage in memory saving compared to the economy-size matrix SVD, even if we recover some left-singular tensors in the full format in order to show the dominant features.  The Hankel singular values have a similar decay pattern to those computed from matrix SVD for smaller truncation thresholds $\epsilon$. The images of the first few left-singular tensors from the two videos quantitatively show that the protein FADD can keep the cancer cell stationary during the cell cycle.

\subsection{Room impulse responses dataset}
The reduction of room reverberation has become increasingly significant in human life. In this example, we try to apply HOERA/ERA to a real world database of binaural room impulse responses (BRIR), referred to as the Aachen Impulse Response (AIR) database \cite{5201259, Jeub10dowe}. The goal is using HOERA to efficiently and accurately capture the acoustic dynamics of a given room, which may potentially help promote the development of algorithms for dereverberation. The BRIR is a one-dimensional output signal, i.e., scalar, so one can simply apply ordinary ERA to the dataset. However, we observe that the generalized Hankel matrix built from the impulse response is huge given the large sampling frequency, and it also contains many extremely small values from the later time points. These suggest that HOERA may be more suitable in estimating the acoustic dynamics.

The AIR database includes many diverse scenarios. In the following studies, we focus the attention on the case where the room impulse responses are collected by a bottom microphone with a bottom-top mock-up phone in an office (see details in \cite{5201259, Jeub10dowe}). We compare the computational times for identifying the acoustic systems using HOERA and ERA with certain numeric accuracies. Let the number of snapshots be $T=L=2^N-1$ for a positive integer $N$. Then we apply  QTTD to the generalized Hankel matrix $\textbf{H}\in\mathbb{R}^{2^N\times 2^N}$ and obtain a tensor train with $J_n=I_n=2$ for $n=1,2,\dots,N$. The remaining steps are followed as discussed in Algorithm \ref{alg:3.4} to find the approximating acoustic model.

\begin{figure*}[h]
\centering
\includegraphics[width=\textwidth]{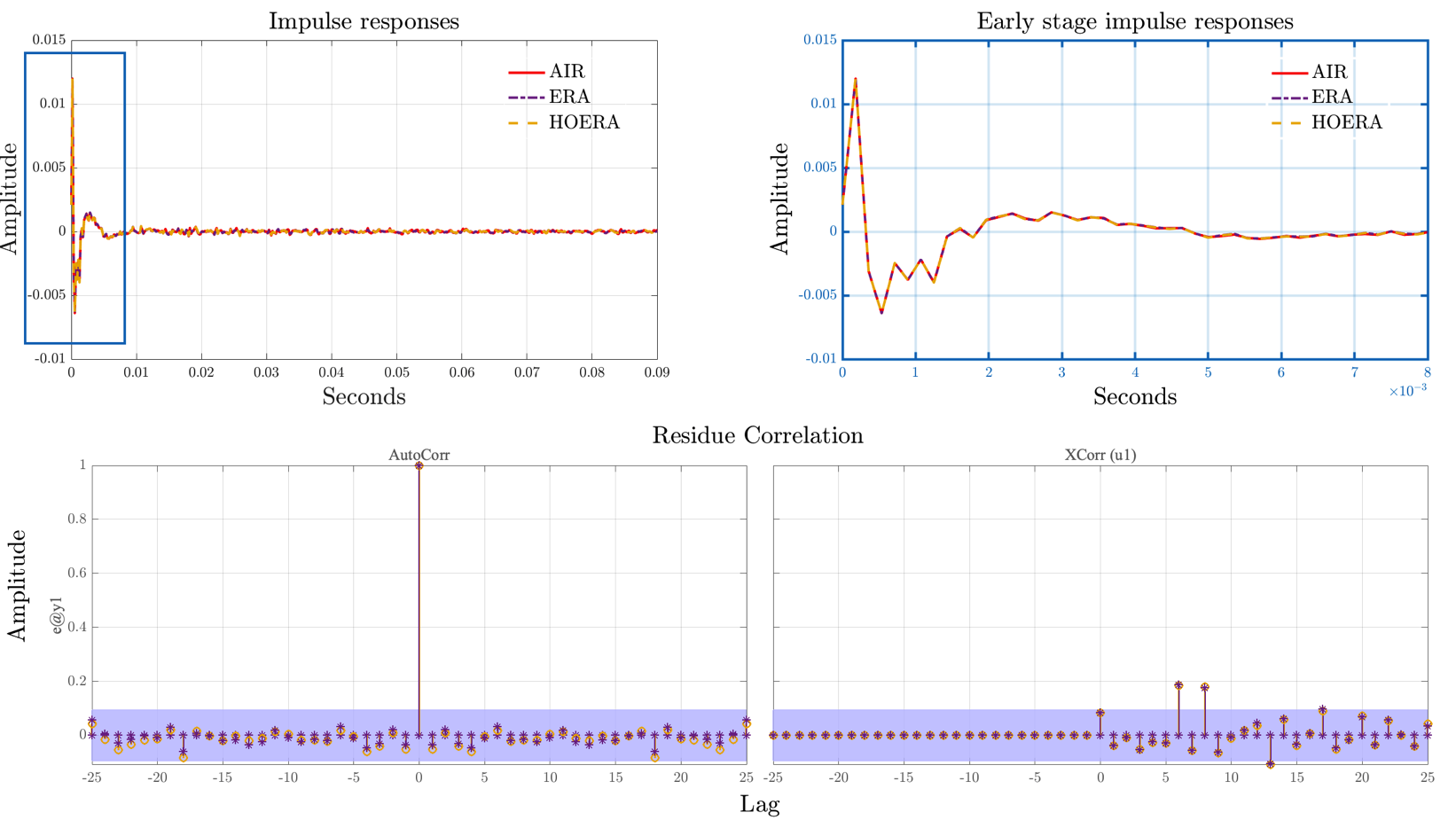}
\caption{The top two figures are the impulse responses from AIR database and models identified by HOERA and ERA, respectively for $N=12$, in which the second one is from magnifying the blue box in the first picture. The bottom figure includes autocorrelation residual and cross-correlation residual plots for ERA and HOERA.}
\label{fig:4.3}
\end{figure*}

The results are given in Table \ref{tab:4.3} and \ref{tab:4.4}.  In the first study, with the increase of $N$, we can allow larger prescribed truncation thresholds in the NPTT-conversion (the truncation in Algorithm \ref{alg:a2}), which accelerates the computation of ETSVD. HOERA exhibits computing time advantage in estimating the system model with low relative errors for $N\geq 12$, see Figure \ref{fig:4.3}. Of course, ERA can provide better relative errors with the same number of singular values retained as in HOERA. In the second study, when $N=13$, larger prescribed truncation thresholds in the NPTT-conversion enable faster computations and lead to similar relative errors compared to ERA, see Table \ref{tab:4.4}. Note that we report the times of computing ETSVD with the conversion time to QTTD and economy-size matrix SVD for HOERA and ERA, respectively, in the both studies.

\begin{table}[h]
\caption{Computational times of HOERA and ERA for different values of $N$ with sampling frequency 5600. ``Threshold'' represents the maximum prescribed truncation threshold required in the NPTT-conversion to maintain the relative errors less than 0.1. ``Relative Error$^{1}$'' and ``Relative Error$^{2}$'' represent the relative errors between the real data and the impulse responses generated by the models identified by HOERA and ERA, respectively.}
{\footnotesize
\begin{center}
\begin{tabular}{|c|c|c|c|c|} \hline
\bf N & $N=9$&  $N=10$& $N=11$ &   $N=12$ \\ \hline
\bf Threshold & 0.9 & 1.2 & 1.5 & 2.4 \\\hline
\bf HOERA (s)  &0.5075 & 2.2262 &  8.2137 & 10.5300  \\ \hline
\bf ERA (s) & 0.1961& 0.9432 &  7.1086 & 50.6742  \\\hline
\bf Relative Error$^{1}$ &0.0929 & 0.0972  & 0.0932 & 0.0849  \\\hline
\bf Relative Error$^{2}$ &0.0890 & 0.0715& 0.0667 & 0.0666 \\
\hline
\end{tabular}
\end{center}
}
\label{tab:4.3}
\end{table}

\begin{table}[h]
\caption{Computational times of HOERA and ERA for different prescribed truncation thresholds with $N=13$ and sampling frequency 12000. ``\# of Singular Values'' represents the minimum number of singular values required to maintain the systems with relative good accuracies for HOERA.}
{\footnotesize
\begin{center}
\begin{tabular}{|c|c|c|c|c|} \hline
 \bf Threshold & $\epsilon=1$&  $\epsilon=2$& $\epsilon=5$ &   $\epsilon=10$ \\ \hline
 \bf \# of Singular Values & 511 & 297 & 72 & 41\\ \hline
\bf HOERA (s)  &198.5078 & 105.8521 &  40.5420 & 17.3069  \\ \hline
\bf ERA (s)  &336.5543 & 336.5543 &  336.5543 & 336.5543  \\ \hline
\bf Relative Error$^1$ &0.0721 & 0.0887  & 0.1086 & 0.1215  \\\hline
\bf Relative Error$^2$ &0.0492 & 0.0835  & 0.1252 & 0.1282  \\\hline
\end{tabular}
\end{center}
}
\label{tab:4.4}
\end{table}

\section{Discussion}\label{sec:5}
The numerical studies reported in the previous section highlight the computational efficiency of different flavors of the proposed HOBT approaches compared to the analogous standard matrix based BT for model reduction/identification of high-dimensional MLTI/LTI systems. These gains come from low rank/sparse structure in the data/underlying system which tensor decomposition methods such as the TTD exploit. While TT-representation and associated algebra are at the core of our proposed framework, it may not always be beneficial to work with the TT-format. In particular, as pointed out in  section \ref{sec:HOBPOD} and \ref{sec:HOERA}, if the number of snapshots, $T$ or $L$, is large, computing the block TT-format and TT-Einstein product can become computational bottlenecks, and it may be preferable to work with the full tensor format or even the unfolded matrix representation. In some circumstances, it may still be possible to control growth of TT-ranks by appropriate truncation, choice of blocking operations and choice of appropriate factorizations used in the block tensor construction as discussed in section \ref{sec:HOBPOD}. From our numerical studies, we found that QTTD is preferable to TTD as it is less sensitive to such choices. Additionally, as proposed in  section \ref{sec:3.1.2}, \ref{sec:HOBPOD} and \ref{sec:HOERA}, one can switch to matrix representation/computation using unfolding at appropriate stages in the algorithm. However, more theoretical and numerical investigation is required to assess under what conditions TT-format versus full tensor format versus matrix representation is computationally more efficient, and is an important avenue of future research. Finally, it should be noted that the proposed framework can also be applied to the standard vector based LTI system representation by applying QTTD directly as we demonstrated in the room impulse response example. 

Furthermore, tensor analysis has played an increasingly significant role in machine learning recently \cite{8187112,WANG201986}. The cancer cell image video example using HOERA/ERA is similar to the eigenfaces problem \cite{139758}, which indicates that the framework of computing ETSVD using TTD would also aid in machine learning algorithms such as principal component analysis, regression analysis and spectral graph theory for large sparse datasets. In particular, ETSVD can help accelerate the computation of von Neumann entropy for large sparse graphs like web graphs and internet topology graphs  for example \cite{sun2007less}.

\section{Conclusion}\label{sec:6}
In this paper, we developed a Higher-Order Balanced Truncation, Higher-Order Balanced Proper Orthogonal Decomposition and Higher-Order Eigensystem Realization Algorithm based model reduction and identification approach for input/output multilinear/linear time invariant systems. The proposed framework exploits tensor decompositions such as tensor train decomposition (including standard TTD, generalized TTD and quantized TTD) and economy-size tensor singular value decomposition to compress the tensors which facilitate efficient computations while retaining the accuracy. We applied our data-deriven model reduction/identification framework to real world biological and engineering systems and achieved outstanding performances in computational efficiency and memory consumption with rigorously quantified errors. We are also currently developing a Dynamic Mode Decomposition (DMD) type variant of our framework. Some initial work generalizing DMD using TTD appears in \cite{Klus_2018}. It would also be worthwhile to develop an observer and feedback control design framework and associated tensor based computational techniques for estimation/control of MLTI/LTI systems and apply these techniques in real world complex systems. Nonlinearity and stochasticity in tensor based dynamical system representation and analysis framework are also important for future research.

% or


\appendices
\section{Supplementary algorithms}\label{appendix}

%\begin{algorithm}[!htbp]
%\caption{Inverse $n$-mode row block tensors}
%\label{alg:a3}
%\begin{algorithmic}[1]
%\STATE{Given $\textsf{C}\in\mathbb{R}^{\mathcal{J}\otimes\mathcal{I}}$ in the GTT-format with cores $\textsf{C}^{(n)}$ and GTT-ranks $\mathcal{R}$, an integer $n$ such that $I_n$ is even}
%\STATE{Set $\textbf{C}^{(n)} = \texttt{reshape}(\textsf{C}^{(n)}, R_{n-1}J_nI_n, R_n)$}
%\STATE{Let $\textsf{A}$ and $\textsf{B}$ be the generalized tensor trains copied \\from $\textsf{C}$ except the $n$-th cores
%\begin{equation*}
%\begin{split}
%\textsf{A}^{(n)} &= \texttt{reshape}(\textbf{C}^{(n)}_{[1:\frac{1}{2}R_{n-1}J_nI_n]:},\\& R_{n-1},J_n,\frac{1}{2}I_n,R_n), \\ \textsf{B}^{(n)} &= \texttt{reshape}(\textbf{C}^{(n)}_{[\frac{1}{2}R_{n-1}J_nI_n+1:R_{n-1}J_nI_n]:},\\& R_{n-1},J_n, \frac{1}{2}I_n, R_n),
%\end{split}
%\end{equation*}
%respectively
%}
%\RETURN $\textsf{A},\textsf{B}\in\mathbb{R}^{\mathcal{J}\otimes\mathcal{L}}$ in the GTT-format with cores $\textsf{A}^{(n)}, \textsf{B}^{(n)}$ and GTT-ranks $\mathcal{R}$, respectively where \\$\mathcal{L} = \mathcal{I}$ except $L_n=\frac{1}{2}I_n$.
%\end{algorithmic}
%\end{algorithm}

\begin{algorithm}[!htbp]
\caption{Conversion of GTTD to TTD}
\label{alg:a1}
\begin{algorithmic}[1]
\STATE{Given $\textsf{A}\in\mathbb{R}^{\mathcal{J}\otimes\mathcal{I}}$ in the GTT-format with cores $\textsf{A}^{(n)}$ and GTT-ranks $\mathcal{R}^g$}\\
\FOR{$n = 1,2,\dots,N$}
\STATE{Set $\textbf{A}=\texttt{reshape}(\textsf{A}^{(n)},R^g_{n-1}J_n, I_nR^g_n)$}
\STATE{Compute  the economy-size SVD of $\textbf{A}$, i.e., $\textbf{A}=\textbf{U}\textbf{S}\textbf{V}^\top$ with $\textbf{S}\in\mathbb{R}^{R_n\times R_n}$ }\\
\STATE{Set $\textsf{X}^{(2n-1)} = \texttt{reshape}(\textbf{U}, R^g_{n-1}, J_n, R_n)$, $\textsf{X}^{(2n)} = \texttt{reshape}(\textbf{S}\textbf{V}^\top, R_n, I_n, R^g_n)$}\\
\ENDFOR
\RETURN $\textsf{A}\in\mathbb{R}^{\mathcal{J}\otimes\mathcal{I}}$ in the TT-format with cores $\textsf{X}^{(n)}$\\ and TT-ranks $\mathcal{R}^{s}=\{R_0^g,R_1,R_1^g\dots,R_{N},R_N^g\}$.\\
\end{algorithmic}
\end{algorithm}

\begin{algorithm}[!htbp]
\caption{Conversion of TTD to NPTTD \cite{tttoolbox}}
\label{alg:a2}
\begin{algorithmic}[1]
\STATE{Given $\textsf{A}\in\mathbb{R}^{\mathcal{J}\otimes\mathcal{I}}$ in the TT-format with cores $\textsf{X}^{(n)}$ \\and TT-ranks $\mathcal{R}^s$}\\
\STATE{Set $\tilde{\textsf{X}}^{(n)} = \textsf{X}^{(n)}$, $\tilde{\mathcal{R}}^s = \mathcal{R}^s$, $Sz=\{J_1,I_1,\dots,$\\$J_N,I_N\}$, $k=p=1$ and $l=2$}
\STATE{Apply right-orthonormalization from the cores $\tilde{\textsf{X}}^{(2N)}$\\ to $\tilde{\textsf{X}}^{(3)}$}\\
\WHILE{$l < 2N$}
\STATE{Apply left-orthonormalization from the cores $\tilde{\textsf{X}}^{(k)}$ \\to $\tilde{\textsf{X}}^{(l-1)}$ for $k\leq l-1$}\\
\STATE{Set $k=l$ and $\textsf{X} = \texttt{reshape}(\bar{\tilde{\textbf{X}}}^{(k)}\underline{\tilde{\textbf{X}}}^{(k)}, \tilde{R}^s_{k-1},Sz_k,Sz_{k+1},\tilde{R}^s_{k+1})$}\\
\STATE{Set $\textsf{X} = \texttt{permute}(\textsf{X}, [1,3,2,4])$ and compute the economy-size matrix SVD of $\textbf{X} = \texttt{reshape}(\textsf{X},\tilde{R}^s_{k-1}Sz_{k+1}, Sz_{k}\tilde{R}^s_{k+1})$, i.e., $\textbf{X} = \textbf{U}\textbf{S}\textbf{V}^\top$}\\
\STATE{Let $\tilde{R}^s_{k} = \text{rank}(\textbf{S})$ and set $\tilde{\textsf{X}}^{(k)} = \texttt{reshape}(\textbf{U}\textbf{S}, \tilde{R}^s_{k-1}, Sz_{k+1}, \tilde{R}^s_{k})$ and $\tilde{\textsf{X}}^{(k+1)} = \texttt{reshape}(\textbf{V}^\top, \tilde{R}^s_{k}, Sz_k, \tilde{R}^s_{k+1})$ }\\
\STATE{Swap the $k$ and $k+1$-th elements in $Sz$ and set $k = \max\{k-1,1\}$}\\
\IF{$k=p$}
\STATE{Set $l = 2p + 2$ and $p = p+1$}
\ELSE
\STATE{Set $l = k$}
\ENDIF
\ENDWHILE
\RETURN $\textsf{A}_{\text{np}}\in\mathbb{R}^{\mathcal{J}\times \mathcal{I}}$ in the TT-format with cores $\tilde{\textsf{X}}^{(n)}$ and TT-ranks $\tilde{\mathcal{R}}^s$.\\
\end{algorithmic}
\end{algorithm}

\section{List of acronyms}\label{apd:B}
\begin{enumerate}
    \item AdCre - Adenovirus Expressing Cre Recombinase 
    \item ADI - Alternating Direction Implicit
    \item AIR - Aachen Impulse Response
    \item BPOD - Balanced Proper Orthogonal Decomposition
    \item BRIR - Binaural Room Impulse Responses
    \item BT - Balanced Truncation
    \item CFL - Courant-Friedrichs-Lewy
    \item CPD - CANDECOMP/PARAFAC Decomposition 
    \item DMD - Dynamics Mode Decomposition 
    \item DMRG - Density Matrix Renormalization Group
    \item ERA - Eigensystem Realization Algorithm 
    \item ETSVD - Economy-Size Tensor Singular Value Decomposition 
    \item FADD - Fas-Associated Death Domain
    \item GTT - Generalized Tensor Train
    \item GTTD - Generalized Tensor Train Decomposition
    \item HOBPOD - Higher-Order Balanced Proper Orthogonal Decomposition
    \item HOBT - Higher-Order Balanced Truncation 
    \item HOEG - Higher-Order Empirical Gramian
    \item HOERA - Higher-Order Eigensystem Realization Algorithm
    \item HOSVD - Higher-Order Singular Value Decomposition 
    \item LTI - Linear Time Invariant 
    \item MLTI - Multilinear Time Invariant
    \item MP - Moore Penrose
    \item NPTT - Non-Paired Tensor Train
    \item NPTTD - Non-Paired Tensor Train Decomposition
    \item QTT - Quantized Tensor Train
    \item QTTD - Quantized Tensor Train Decomposition 
    \item SVD - Singular Value Decomposition
    \item TSVD - Tensor Singular Value Decomposition 
    \item TT - Tensor Train 
    \item TTD - Tensor Train Decomposition 
    
\end{enumerate}

\section*{Acknowledgment}
We thank Dr. Frederick Leve at the Air Force Office of Scientific Research (AFOSR) for support and encouragement. This work is supported in part under AFOSR Award No: FA9550-18-1-0028, NSF grant DMS 1613819, Smale Institute, and the Lifelong Learning Machines program from DARPA/MTO.


\ifCLASSOPTIONcaptionsoff
  \newpage
\fi

\end{document}